\documentstyle[aps]{revtex}
\newcommand{\lettersize}{}
\newcommand{\dir}{FIGS}
\input psfig
\input{psfig}
\newcommand{\fig}[3]
{
\begin{center}
     \noindent
     \unitlength=1mm
     \begin{picture}(#2,#3)
     \put(0,0){
       \psfig{figure=\dir/#1,width=#2mm,height=#3mm}
     }
     \end{picture}
   \noindent
\end{center}
}
\begin{document}
\baselineskip=12pt
\setcounter{page}{1}

\noindent
{\LARGE\bf
``Intrinsic'' profiles and capillary waves at homopolymer interfaces:
a Monte Carlo study
}

\vspace{0.5cm}

\lettersize

\begin{center}
A. Werner, F. Schmid, M. M\"uller, K. Binder \\
{\em Institut f\"ur Physik, Universit\"at Mainz, D-55099 Mainz, Germany}
\end{center}


\begin{quote}
{\bf Abstract.}
A popular concept which describes the structure of polymer interfaces
by ``intrinsic profiles'' centered around a two dimensional surface,
the ``local interface position'', is tested by extensive Monte Carlo
simulations of interfaces between demixed homopolymer phases in symmetric
binary (AB) homopolymer blends, using the bond fluctuation model. The
simulations are done in an $L \times L \times D$ geometry. The interface
is forced to run parallel to the $L \times L$ planes by imposing periodic
boundary conditions in these directions and fixed boundary conditions in
the $D$ direction, with one side favoring A and the other side favoring B.
Intrinsic profiles are calculated as a function of the
``coarse graining length'' $B$ by splitting the system into columns of
size $B \times B \times D$ and averaging in each column over profiles 
relative to the local interface position. The results are compared
to predictions of the self-consistent field theory.
It is shown that the coarse graining length can be chosen such that the 
interfacial width matches that of the self-consistent field profiles, and 
that for this choice of $B$ the ``intrinsic'' profiles compare well with the
theoretical predictions. Our simulation data suggest that this 
``optimal'' coarse graining length $B_0$ exhibits a dependence of the form 
$B_0 = 3.8 \: w_{\rm SCF} (1 - 3.1/\chi N)$, where $w_{\rm SCF}$ is
the interfacial width, $N$ the chain length and $\chi$ the Flory-Huggins
parameter.
\end{quote}

\lettersize

\section{Introduction}

Polymer blends\cite{blends,blends2} are examples of systems which can usually 
be described very well by mean field theories: Due to the chain connectivity,
the effective range of interactions 
between polymers, which is roughly the extension of the chains, becomes very 
large for high molecular weights, and according to the Ginzburg criterion, the
critical region in which critical fluctuations become important is
very small as a result\cite{ginzburg}. 
Hence fluctuations can usually be neglected except in the ultimate vicinity 
of the critical point.

If interfaces are present, however, there exists a type of 
fluctuations which survives even deep in the two phase region. 
This is because the interface breaks a continuous symmetry, 
the translational invariance. As a consequence, long wavelength
transversal excitations come into existence (Goldstone bosons).
The  energy of these ``capillary waves'' of the local interface 
position\cite{capillary} vanishes as the wavelength approaches infinity.
These fluctuations are not taken into account in mean field approximations.
Nevertheless, they strongly influence all quantities which depend on
transversal degrees of freedom. 

For example, the interfacial waves
contribute to the total width of the interface in a way that it diverges
logarithmically with the lateral system size\cite{capillary}. In other words, 
the apparent width of the interface depends on the length scale on which the 
interface is studied. This important observation is not just of academic
interest, but also relevant for technical reasons: 
The mechanical stability of interfaces is to a great extent determined by 
the number of entanglements between polymers of different type which
is, in turn, closely related to the interfacial width
on the length scale of at most the radius of gyration $R_g$. 
On the other hand, experiments which measure the interfacial width
usually work with lateral resolutions characterized by much larger
length scales. Hence, the experimental results cannot be directly 
related to  the mechanical properties, and a careful analysis
of capillary wave effects is required\cite{sem1,shull}.

Despite their success in the description of general bulk thermodynamics,
mean field approaches thus apparently fail to capture essential
properties of polymer interfaces. However, the situation is not entirely
hopeless. Capillary waves are the only Goldstone bosons present in the
system, and one may safely assume that they are the only fluctuations
which remain important outside of the critical region. Further
simplification can be achieved by neglecting the coupling of long
wavelength capillary wave fluctuations to the local interfacial
structure. This leads to a simple picture in which the
interface is described by:\cite{capillary}
\begin{itemize}
\item the local interface position, a function $h$ which parameterizes
a two dimensional surface and which is distributed according to a
capillary wave Hamiltonian ${\cal H}\{ h\}$;
\item local intrinsic profiles, which are centered around the local
interface position, but do not depend on $h$ otherwise, and which can
be calculated within an appropriate mean field theory.
\end{itemize}
The intrinsic profiles characterize the 
interface on a certain length scale which has yet to be specified, 
{\em i.e.}, the coarse graining length.  
The theory thus assumes that one can define a coarse graining length on 
which mean field theory provides a valid description of the interfacial 
structure. If this is the case, it should be related to one of the
natural length scales in the system, {\em i.e.}, the radius of gyration
of the chains, the intrinsic width of the interface, or some
microscopic length such as the monomer size.

We shall quantify this picture in more detail for the special case of a 
planar interface. Neglecting bubbles and overhangs, the local interface 
position can then be parameterized by a single-valued function $h(x,y)$. 
Long wavelength capillary fluctuations basically cost the free energy
associated with the increase of interfacial area. Hence the capillary
wave Hamiltonian is given by\cite{sem1,kerle,sferraza}
\begin{equation}
\label{CW}
{\cal H}_{CW}\{h\} = 
\sigma \int \! dx \, dy \, \Big\{ 
\sqrt{1+\big({\partial h}/{\partial x}\big)^2} 
\sqrt{1+\big({\partial h}/{\partial y}\big)^2} - 1 \Big\}
\approx \frac{\sigma}{2} \int \! dx \, dy | \nabla h |^2,
\end{equation}
where $\sigma$ is the interfacial tension, and $|\nabla h| \ll 1$ has
been assumed. Since it is essentially quadratic in $h$, thermal averages
can be carried out easily. The functional ${\cal H}_{CW}$ can be
diagonalized by means of a Fourier transformation with respect to $x$ and $y$,
resulting in 
\mbox{${\cal H}_{CW} \{ h\} = \sigma/2 \: \sum q^2 |h(\vec{q})|^2$},
and the thermal average of $h(\vec{q})$ takes the value
\begin{equation}
\label{hq}
\langle |h(\vec{q})|^2\rangle = 1/(\sigma q^2).
\end{equation}
For the local interface position, one finds a Gaussian height distribution
\begin{equation}
\label{ph}
P_L(h) =
1/\sqrt{2 \pi s^2} \exp(-h^2/2 s^2)
\end{equation}
with
\begin{displaymath}
s^2 = \frac{1}{4 \pi^2} \int d \vec{q} \langle |h(\vec{q})|^2 \rangle
= \frac{1}{2 \pi \sigma} \ln (\frac{L}{B_0}).
\end{displaymath}
Here $(2 \pi/L)$ and $(2 \pi/B_0)$ had to be introduced as lower and upper
cutoff of the integral 
\mbox{$\int d \vec{q} \langle | h(\vec{q})|^2 \rangle \sim \int dq/q$},
which diverges both for $q \to 0$ and for $q \to \infty$. The large
length $L$ is set by the system size, and the small length $B_0$ is
the coarse graining length mentioned above on which the interface
assumes its ``intrinsic'' structure. 

The intrinsic profiles can be calculated using one of the various 
sophisticated mean field approaches which have been specifically
designed to study inhomogeneous polymer 
systems\cite{helfand,flory,freed,schweizer}.
In our work, we use Helfand's self-consistent field theory, which
treats the polymer chains as random walks in the self-consistent field
created by the surrounding polymers\cite{helfand,friederike1}.
Given the intrinsic density profile $\rho_{Q,intr}(z)$ of a given 
quantity $Q$, the general profile $\rho_Q(x,y,z)$ takes the form
$\rho_Q(x,y,z) = \rho_{Q,intr}(z-h(x,y))$, 
and after performing the thermal average over the capillary wave
fluctuations, one obtains the ``apparent'' profiles (convolution approximation):
\begin{equation}
\label{app}
\rho_{Q,L}(z) = \int \! dh \, \rho_{Q,intr.}(z-h) \: P_L(h).
\end{equation}
For example, the interfacial width which we define as
\begin{equation}
\label{width}
w := \rho \Big/ \: \Big|\frac{d (\rho_A-\rho_B)}{dz}\Big|_{z = h},
\end{equation}
in a binary (AB) blend, is broadened according to\cite{sem1,andreas2}
\begin{equation}
\label{wl}
w_L^2 = w_{intr}^2 + \frac{1}{4 \sigma} \ln(\frac{L}{B}).
\end{equation}
For polymer interfaces, this relation was originally derived by
Semenov\cite{sem1}, who suggested that the coarse graining length
is given by $B = \pi w_{intr}$.

The use of a sharp cutoff $B$
in the capillary Hamiltonian is of course not a rigorous procedure.
Compared to the length scale $B$ the intrinsic width is not
necessarily small, and, hence, it is possible that bulk fluctuations and
fluctuations of the local interface position become coupled on this
small scale. Furthermore, higher order terms, such as $|\Delta h|^2$,
could become important. Thus, a test of the accuracy of the present approach
is clearly warranted.

In the present paper, we aim to provide a detailed test of this picture. 
To this end, we have performed extensive Monte Carlo simulations of 
interfaces between homopolymer phases in a symmetric binary (AB)
homopolymer mixture, using the bond fluctuation model\cite{ck,binder,marcusr}.
The presence of capillary waves at polymer interfaces has been
demonstrated experimentally\cite{kerle,sferraza} 
and in Monte Carlo simulations\cite{andreas2,marcus3,gary}.
In an earlier study, we have established quantitatively the validity
of capillary wave concepts for interfaces confined in thin films,
and for effectively free interfaces, {\em i.e.}, in the limit of 
large film thickness $D$\cite{andreas2}. 
Interfacial profiles of various quantities have also been obtained
previously from simulations\cite{marcus1} and compared to mean field 
predictions\cite{friederike2,andreas2,marcus2}.

In two of these studies\cite{andreas2,marcus2}, capillary waves on 
length scales down to a given coarse graining length $B_0$ were
subtracted using a coarse graining procedure.
However, a direct comparison between self-consistent field calculations and 
Monte Carlo results or experiments obviously faces the problem  that the 
coarse graining length of the intrinsic profiles is as yet not known.
In the previous work\cite{andreas2,marcus2}, it was chosen somewhat
{\em ad hoc}, such that the measured interfacial width compares well with 
the calculated width. The way how this ``optimal choice'' of the coarse 
graining length depends on the model parameters, {\em i.e.}, on the various 
natural length scales in the system, was not investigated.  

The present work therefore attempts a systematic study of this coarse graining 
length, in particular its dependence on the chain length and the chain 
incompatibility. To this end,
we have calculated profiles as a function of the coarse graining
length, and performed systematic variations of both the chain length $N$
and the Flory-Huggins parameter $\chi$. As we shall see, our results
can be brought into agreement with the self-consistent field profiles when
using a coarse graining length which scales roughly like 
$B \propto w_{\rm SCF}[1-3.1/(\chi N)]$,
where $w_{\rm SCF}$ is the width of the self-consistent field profile.
This is one of the main results of this work. 
On the other hand, intrinsic profiles are also interesting in their
own right. We shall see for the case of density profiles for contacts
within chains and between chains (contact numbers), that the intrinsic 
profiles may actually differ {qualitatively} from apparent profiles.

Our paper is organized as follows: In the next section, we introduce
the bond fluctuation model and the simulation technique and comment
briefly on the self-consistent field calculations. In section III, we
analyze the capillary wave spectrum of the interface and demonstrate how
it can be used in different ways to extract the interfacial 
tension $\sigma$\cite{marcus2}. Section IV is then devoted to the discussion
of various intrinsic profiles and of the coarse graining length. 
We summarize and conclude in section V.

\section{Model and technical details}

The bond fluctuation model is a refined lattice model for polymer
fluids which has the advantage of combining the computational
efficiency of a lattice model with high versatility,
such that the actual structure of the fluid shows almost no signature of the
structure of the underlying lattice. 
Polymers are modeled as chains of effective monomers, each
occupying a cube (eight sites) of a simple cubic lattice, 
and these monomers are connected by bonds of variable length 
of 2, $\sqrt{5}, \sqrt{6}, 3$, or $\sqrt{10}$ lattice constants. 
At a volume fraction $0.5$ or a monomer number density $1/16$, the polymer 
fluid exhibits the characteristic properties of a dense melt, {\em i.e.}, 
polymer conformations have almost ideal Gaussian chain statistics\cite{pbhk}. 
We consider homopolymers made of two different types of monomers --
A and B -- which interact {\em via} a symmetric potential
 $ \epsilon_{AA}=\epsilon_{BB}=-\epsilon_{AB}= - k_B T \epsilon$
if they are less than $\sqrt{6}$ lattice constants apart from each other
({\em i.e.}, the interaction shell includes 54 neighbor sites). 
Most simulations were done using polymers of length $N=32$; the
$A$ and $B$ homopolymers then demix at $\epsilon > \epsilon_c \approx 0.014$.
However, we increase the chain length up to $N=256$ in order to investigate
the  chain length dependence of the coarse graining length $B_0$.
A well-defined interface is enforced in the canonical ensemble in a thick film 
geometry ($L \times L \times D$), with periodic boundary conditions
in the $L$ directions and walls which favor A on one side and B
on the other side. The wall interacts with the monomers in the first
two layers near the wall, and the interactions were chosen large enough
that the walls are wetted by their favorite phase\cite{marcus4}
({\em e.g.}, for $N=32$ and $\epsilon=0.03$, we choose $\epsilon_w=0.1 k_B T$).
These boundary conditions ensure that the interface is on average located in 
the middle of the film. The film thickness ($D=64$ or 128) is large enough 
compared to $L$ ($L=128$) that the capillary wave fluctuations
are limited by the system size rather than by the film thickness\cite{andreas2},
and that the interactions of the interface with the walls are negligible.
Hence the interface is basically free. 
We equilibrate and sample the system using a combination of local monomer 
moves\cite{ck}, slithering snake moves\cite{kb}, and particle exchange 
moves. The autocorrelation time in the simulations will be 
discussed in section III.

In order to analyze the interfacial fluctuations and intrinsic profiles,
we split the system into columns of block size $B \times B$ and height
$D$ (see Fig. 1) and determine the Gibbs dividing surface 
$h(x,y)$ in each column\cite{andreas2}. This is done by counting the
number of $A$ monomers $n_A$ and of $B$ monomers $n_B$ in the column,
and defining $h = N_A \; D/(N_A+N_B)$.
Profiles of various quantities are then
taken relative to this position. The block size $B$ was varied 
to allow for a systematic analysis.

The results are compared to self-consistent field predictions within
a simple Helfand type theory. In this approach, 
the polymers are described as random walks with statistical weight
\begin{equation}
\label{gauss}
{\cal P}\{ \vec{r}(\cdot)\} = {\cal N}
\exp \big[ -\frac{3}{2 b^2} \int_0^N ds \big|\frac{d \vec{r}}{ds}\big|^2 \big]
\end{equation}
in an external field which is created by a monomer interaction potential
\begin{equation}
\label{helfand}
\beta {\cal F} = \frac{1}{\rho_b}
\int d\vec{r} \: \big\{
\chi \rho_A(\vec{r}) \rho_B(\vec{r}) + \frac{\zeta}{2} (\rho_A+\rho_B-\rho_b)^2
\big\}
\end{equation}
with the monomer bulk density $\rho_b$. The basic parameters of the
model are the statistical segment length $b$ which characterizes
the random walk statistics of the chain, the Flory-Huggins parameter
$\chi$ which describes the relative repulsion between unlike monomers,
and the inverse compressibility $\zeta = 1/(\rho_b k_B T \kappa)$. 
All these parameters can in principle be determined by independent
bulk simulations. The statistical segment length is related to
the radius of gyration of the chains, $R_g = b \sqrt{(N-1)/6}$. It depends 
weakly on the chain length; at fixed $\epsilon N = 0.96$,
we find $b(N) = 3.20(5) - 0.8(2)/\sqrt{N}$, and at fixed $\epsilon = 0.03$,
$b(N) = 3.11(1) - 0.2(1)/\sqrt{N}$. At chain length $N=32$ in particular,
the statistical segment length takes the value $b=3.06$.
The Flory Huggins parameter depends on the monomer interaction parameter
$\epsilon$ and on the average number of interchain contacts of a
monomer,\cite{marcus5} the ``effective coordination number'' $z_{\rm eff}$
\begin{equation}
\label{zeff}
\chi = 2 z_{\rm eff} \epsilon .
\end{equation}
The latter decreases with increasing chain length due to the effect
of the correlation hole\cite{schweizer,degennes}. At fixed $\epsilon N = 0.96$, 
our data can be described by the law $z_{\rm eff}(N) = 2.12(1) - 2.97(6)/\sqrt{N}$
which is comparable to the behavior in the athermal system
($\epsilon = 0$):  $z_{\rm eff}(N) = 2.1 - 2.8/\sqrt{N}$\cite{marcus6}.
Finally, the inverse compressibility has been estimated in an athermal 
system from the entropy density $s$, $\zeta = 4.1$ \cite{marcus6}.
The detailed independent knowledge of the system parameters allows us to compare the
simulational results the self-consistent field calculations without adjustable parameter.

\section{Capillary waves}

This section shall be concerned with the analysis of the pure capillary 
wave spectrum, not bothering yet with intrinsic profiles and coarse
graining lengths. A somewhat similar study has already been presented 
earlier by us\cite{andreas2}, hence we shall be brief for the most part.
Our analysis is needed here to put our later results into context.
In addition, we shall also discuss how the capillary waves can be
exploited in different ways to extract the interfacial tension.

The effect of capillary waves on the apparent profiles of the order
parameter $m(z)=[\rho_A(z)-\rho_B(z)]/\rho(z)$ is demonstrated in
Fig. 2 for different system sizes $L$. One clearly recognizes how
the interface broadens with increasing $L$. Note that the
relaxation time of the capillary waves also grows with the system size
and becomes very large, since the forces which drive the capillary
waves back are very small for long wavelengths. It is crucial
to ensure that the total length of the simulation run is longer than
the time scale which governs the dynamics of the slowest capillary mode. 
In order to check this, we have calculated for each configuration
the Fourier modes\cite{marcus3} of the local interface position function $h(x,y)$ 
in one direction, $h_i = h(\vec{q}_i)$ with $\vec{q}_i = (i,0) \cdot 2 \pi/L$.
The slowest Fourier mode is the lowest mode with $i=1$. Thus the quantity
of interest is the decay time $\tau$ of the corresponding
autocorrelation function
\begin{equation}
C_{hh}(t) =
\frac{\langle h_1(t) h_1(0)\rangle - \langle h_1 \rangle^2}
{\langle h_1{}^2\rangle - \langle h_1 \rangle^2}
\propto \exp(-t/\tau).
\end{equation} 
It is shown in Fig. 3 for the lateral dimension $L=128$
as a function of $\epsilon$. The length of the simulation runs
was generally between $10^6$ and $10^7$ Monte Carlo steps. 
This is much longer than the autocorrelation time $\tau$ for
small $\epsilon$, but gets close to $\tau$ for $\epsilon=0.1$.
Hence, the results for large $\epsilon$ have to be interpreted with
some caution.

From the capillary wave spectrum, one can now calculate the interfacial
tension following three different strategies:
\begin{itemize}
\item[(a)] Direct inspection of $h_i{}^2$ as a function of $(1/i)^2$
(Fig. 4a) and use of eqn. (\ref{hq}).
\item[(b)] Determination of the width $s$ of the local height distribution
function $P_L(h)$ (Fig. 4b) and use of eqn. (\ref{ph}).
\item[(c)] 
Calculation of the apparent interfacial width $w_L$ as a function of
system size $L$ (Fig. 4c) and use of eqn (\ref{wl}).
\end{itemize}

Alternatively to (c), one can also simulate a single (very large) system size 
$L$, perform the block analysis described in section II, and study $w_B$ as a 
function of the block size $B$. This is shown in Fig. 4d. 
For very small $B$, local concentration fluctuations become important
and the description by eqn. (\ref{wl}) no longer applies. At large enough
$B$, however, one observes a nice logarithmic dependence, from which
only the last point at $B=L$ deviates slightly. The latter can be 
understood from the fact that the number of capillary modes contributing
to the broadening of $w_B$ at $B=L$ is reduced by two compared to that
in a larger system $(L>B)$, due to the constraint of periodic boundary
conditions.\cite{andreas2}

The interfacial tensions obtained with these three different methods are
compared with each other and with the theoretical predictions in Fig. 5.
The agreement is very good for small values of the Flory-Huggins parameter $\chi$,
$\chi <0.4$ or $\epsilon <0.07$. At larger incompatibilities, the data
scatter very much due to the fact that the order of magnitude of the
relaxation time $\tau$ (Fig. 3) gets close to the total length of
the simulation runs ($5 \cdot 10^6$ Monte Carlo steps).
In the same regime, the values of the interfacial tensions derived according
to (a), (b), and (c) differ systematically from each other: Those obtained
by (b) are lowest, followed by (a) and (c). 
We cannot excluded that the strategies (a)--(c) yield truly
different values for the interfacial tension at
high incompatibility $\chi$. Much longer runs would be needed
to settle the question whether the deviations are systematic or due to
the protracted long correlation times.
Systematic deviations are also found close to the critical demixing point,
at $\chi < 0.15$ or $\epsilon < 0.03$. Here, critical fluctuations come into 
play, and the capillary wave description of eqn. (\ref{CW}) does not apply 
any more on the length scales of the simulation.
For comparison, Fig. 5 also shows values of the interfacial tension
which have been calculated earlier by some of us using histogram
reweighting techniques\cite{marcus1}. Within the statistical error,
these independent data agree well with the ones obtained here.
The agreement with the theoretical prediction of the self-consistent field 
calculation is also quite good, especially when the interfacial tension
is determined according to (b).

\section{Intrinsic profiles}

\subsection{Density profiles and local compressibility}

We now turn to the discussion of interfacial profiles. 
As already emphasized in the introduction, these generally depend strongly
on the choice of the coarse graining length or block size $B$. 
However, some properties of the interface can also be discussed 
independently of the coarse graining length: In the approximation (\ref{app}),
capillary waves do not affect the total excess of quantities.  
This holds in particular for the total density $\rho = \rho_A + \rho_B$:
Hence, the total mass reduction in the interfacial area should not depend
on the coarse graining length.

To illustrate this, Fig. 6 shows the total density profiles obtained from
coarse graining over blocks of different size $B=8$ or $B=L=128$.
The profiles broaden for larger $B$, but the depth of the density dip
decreases in turn, and the total area remains constant. Also shown
is the prediction of the self-consistent field theory for different values
of the inverse compressibility $\zeta$. The bulk compressibility of the 
melt has been determined in earlier work, leading to $\zeta=4.1$\cite{marcus6}.
However, Fig. 6 indicates that the theoretical profiles calculated with 
this value are not compatible with the simulation data.
Good agreement is reached with $\zeta = 1.9$, {\em i.e.}, assuming
a ``local compressibility'' at the interface which is more than twice
as high than in the bulk. Interestingly, this value of $\zeta$ seems
to lead to a good description of the simulation data { independent} of
the Flory Huggins parameter $\chi$ and the chain length $N$. This
is demonstrated in Fig. 7 for a wide range of monomer interactions
$\epsilon$ (Fig. 7a) and chain lengths $N$ (Fig. 7b).

In order to further quantify this finding, we plot the depth of the dip
in the density profiles at block size $B=8$ as a function of $\epsilon$
in Fig. 8 and compare it with the theoretical prediction for
$\zeta=1.9$ and $\zeta=4.1$. The block size $B=8$ was used
because the overall shape of the profiles is best fitted by the
theory for this choice (cf. section IV.B and Fig. 7).
The theoretical prediction for $\zeta=1.9$ agrees extremely well with the
simulation data, even the deviations from the straight line at small
$\epsilon$ are reproduced quantitatively. In contrast, the prediction
for $\zeta=4.1$ does not fit the data at all.

Hence, it appears that the local density variations at the interface
are governed by a local compressibility which differs significantly from
the overall compressibility of the melt. The reason for this unexpected
finding is not clear. Previous simulation studies of local bulk density
fluctuations in a similar system have rather suggested that the 
compressibility should slightly increase on short length scales\cite{joerg}.
This question will clearly need to be investigated further in future studies.

\subsection{Intrinsic width and ``coarse graining length''}

Next we consider the density profile $\rho_A(z)$ of a single component
which is much more fundamentally affected by the capillary waves. 
As explained in section II, it is used to locate the interface position
and to define the interfacial width $w$ (eqn. (\ref{width})).
Fig. 4d shows the interfacial width as a function of block size $B$
for various $\epsilon$ and fixed chain length $N=32$, also indicating
the predictions of the self-consistent field theory. For all values
of $\epsilon$, the theoretical prediction and the simulation results agree
best with each other if the block size $B$ is chosen $B=8$.
This is demonstrated even more convincingly in Fig. 9 which 
compares the interfacial width for block size $B=8$ and $B=L=128$
with the self-consistent field results ($\zeta=1.9$ and $\zeta=4.1$)
over a wide range of Flory-Huggins parameter $\chi$.
Except very close to the critical point, the quantitative agreement
for $B=8$ is very good. Note that the theoretical curves for the
interfacial width do not depend very strongly on the inverse compressibility
$\zeta$. The most notable compressibility effect is observed when
plotting $w$ in units of $w_{SSL}=b/\sqrt{6 \chi}$. 
In an incompressible blend, $w/w_{SSL}$ should approach one smoothly from
above as $\chi$ increases. In a compressible system, it first decreases,
reaches a minimum and then rises again. This is found consistently
both in the simulations and in the self-consistent field calculations
for $\zeta=1.9$.

We turn to the discussion of the optimal choice of the block size $B$,
which is the coarse graining length for the intrinsic profiles
mentioned in the introduction. As discussed there, it should be
related to some natural length scale of the system, {\em i.e.},
some microscopic length like the monomer size or statistical segment
length, the intrinsic width, or the radius of gyration of the chains.
In the first case, it should be independent of both $\epsilon$ and $N$.
In the second case, it should depend on $\epsilon$. So far, we have seen
no indication of such a dependence -- however, the interfacial width
varies so little with $\epsilon$ in the range considered by us 
(Fig. 9) that we cannot draw any conclusions from this observation.
In the last case, it should scale like $\sqrt{N}$ with the chain length $N$. 

In order to test the different possibilities, we have performed simulations for 
chain lengths $N$ of up to $256$ , keeping either $\epsilon$ constant 
($\epsilon=0.03$) or $\epsilon N$ constant ($\epsilon N = 0.96$). 
The resulting interfacial width is shown as a function of block size and
compared with the self-consistent field prediction in Fig. 10. 
One finds that the optimal block size $B_0$ now differs for the different 
parameters and depends on the chain length $N$.

The results are summarized in Fig. 11. At fixed $\epsilon N=0.96$,
{\em i.e.}, at constant $\chi N$, the optimal block size $B_0$ scales 
like $B_0 \propto \sqrt{N} \propto 1/\sqrt{\chi}$ (Fig. 11a, inset).
This rules out the possibility that $B_0$ should be related to some
microscopic length scale, and establishes a relationship
of the form $B_0 = R_g f(\chi N) = w_{\rm SCF} \tilde{f}(\chi N)$. (Note that
$w_{\rm SCF} \propto R_g/\sqrt{\chi N} \big[ 1-\alpha/(\chi N)+ 
\cdots \big]^{-1/2}$
with $\alpha \approx 2.5$\cite{broseta,freed2}).
At fixed $\epsilon = 0.03$, {\em i.e.}, at constant $\chi$, the optimal
block size $B_0$ first increases with $N$, but levels off faster than 
$\sqrt{N}$ at the largest chain length $N=256$ (Fig. 11a). 
The data seem to approach a constant in the strong segregation
limit $N\to \infty$. This long chain length behavior becomes even clearer 
when plotting $B_0/w_{\rm SCF}$ as a function $1/(\chi N)$ (Fig. 11b). 
The data can be described satisfactorily by the function
\begin{equation}
\label{b}
B_0 = 3.8 \; w_{\rm SCF} \; \big(1-3.1 \; (\chi N)^{-1} \big),
\end{equation}
and all three data sets -- those for variable chain length $N$ at fixed 
incompatibility $\epsilon$ (circles), those for variable incompatibility at 
fixed chain length (stars), and those where the chain length and 
incompatibility have been varied such that the product $\epsilon N$ remains 
constant -- collapse onto this single master curve. 
Hence, our simulational results suggest that the coarse graining length is a 
multiple of the intrinsic width subjected to strong chain end corrections of 
order $1/(\chi N)$. Note that the pronounced chain end effects are rather 
unexpected, because we have used the self-consistent field result 
in eqn. {\ref b}, which already include a correction of a similar form.

Upon increasing the incompatibility at fixed chain length, the interfacial 
width $w_{\rm SCF}$ decreases, and the chain end correction factor to $B_0$ 
increases. Therefore the actual value of the optimal block size $B$ for chain 
length $N=32$ has a a maximum at $\chi \approx 0.2$ and varies very little 
($B \approx 7$) in the range of $\chi$ considered in the previous sections, 
$\chi \in [0.1:0.8]$. This explains why such good results were obtained 
with constant block size $B=8$.

Hence, we have shown that our simulation data can be analyzed consistently
within the concepts which we have developed in the introduction, {\em i.e.}, 
assuming the existence of intrinsic profiles which can be obtained
from mean field theory and are broadened by capillary waves.

\subsection{Intrinsic profiles of other quantities}

In this last section, we shall discuss selected profiles of other
quantities and relate them to self-consistent field predictions. 
We restrict ourselves to chain length $N=32$ and 
calculate the intrinsic profiles by coarse graining over blocks of
block size $B=8$ for a broad range of incompatibilities.

Fig. 12 shows profiles of the relative density of chain ends. They
enrich at the interface for entropic reasons. This, in turn, creates
a depletion zone at a distance of a radius of gyration from the 
center of the interface. The height of the peak is
slightly underestimated by the theory, yet the overall agreement
is still good.

Next we consider the orientational properties of chains. 
Polymers generally tend to orient themselves parallel to surfaces
and interfaces. Two different factors are involved in
this behavior: Reorientation of chains without distortion of the intrinsic 
shape, {\em i.e.},
at constant absolute value of the radius of gyration or end-to-end vector,
and chain compression towards the interface. 
The self-consistent field theory for Gaussian chains can only handle
chain compression, since the $x$, $y$ and $z$ directions are decoupled
in random walks ({\em i.e.}, the $x$ and $y$ components $R_{ee,x}{}^2$ and
$R_{ee,y}{}^2$ are not affected by the presence of the interface.)
In our simulations, both effects are present, yet chain compression
is by far dominant (Fig. 13 a). The $z$ component of the end-to-end
vector is reduced to almost 30 \% at the interface.
The profiles of $R_{ee,z}{}^2$ are very well reproduced by the 
self-consistent field calculations, even in details such as the
slight overshoot at distances from the interface of about two radii of 
gyration. The agreement is not quite as good when looking separately at 
the orientation of chains which are in their minority phase. Close to 
the interface, $R_{ee,x}{}^2$ and $R_{ee,y}{}^2$ are then found to 
increase by up to 50 \% which indicates that chain reorientation takes 
place and that the chains are even somewhat stretched parallel to the 
interface. Deep in the bulk, the total dimensions of minority chains are 
reduced compared to those of majority chains (Fig. 13b, cf. Sariban
and Binder\cite{sb0}).

Another quantity of interest is the bond orientational parameter $q_b$
which characterizes the orientation of single bonds,
\begin{equation}
q_b = \frac{ \langle l_z{}^2 \rangle 
- (\langle l_x{}^2 \rangle + \langle l_y{}^2 \rangle)/2}
{\langle \vec{l}^2 \rangle },
\end{equation}
where $\vec{l}$ are the bond vectors. The bond orientational parameter is
positive for perpendicular orientation and negative for parallel 
orientation. Fig. 14 shows profiles of $q_b$ for various values of
the monomer interaction strength $\epsilon$. Like whole chains,
single segments also orient parallel to the interface, but to much
less extent\cite{marcus1}.
Segment orientations are not accessible to self-consistent field studies
of Gaussian chains since random walks do not have well-defined tangent
vectors. In order to calculate them, one has to resort to a different
chain model, {\em e.g.}, the wormlike chain model which describes chains
as strings of fixed contour length with a conformational weight functional
governed by a bending rigidity $\eta$\cite{KP,mf}. The latter is related
to the statistical segment length {\em via} $b = \sqrt{2 \eta} \; a$, where
$a$ is the monomer length. In the case of the bond fluctuation model,
subsequent bonds are essentially uncorrelated except for the fact
that they cannot fold back onto themselves. Hence $b \approx a$ or
$\eta \approx 1/2$ seems like a reasonable guess for the effective
bending stiffness. Bond orientational profiles have been calculated
earlier within the wormlike chain model for $\epsilon=0.1$ and various
values of $\eta$\cite{friederike2}. At $\eta=0.5$, the minimum of $q_b$ at the
center of the interface takes the value $q_b=-0.025$, {\em i.e.}, it 
underestimates the simulations by a factor of two. Better agreement
for all $\epsilon$ is reached with $\eta=1.2$. However, such a
high bending stiffness would imply that the monomer length is unreasonably
small, $a=1.97$ which is smaller than smallest possible bond length 2
in the bond fluctuation model. The average bond length
is $\sqrt{\langle l \rangle^2}=2.62$. 
Thus one is lead to suspect that the wormlike chain model does not
describe the chains of the bond fluctuation model any better than
the Gaussian chain model. Note that the wormlike chain model reduces
to the Gaussian chain model in the limit $\eta \to 0$\cite{mf}.
However, taking account of the detailed chain architecture in the mean 
field framework better agreement could be achieved\cite{marcus2}.

Finally, we shall examine the profiles of the average contact number density
for contacts between monomers of different chains $N_{inter}$ and between 
monomers of the same chain $N_{intra}$. One of the fundamental
assumptions of the mean field theory is that they should behave like
$\rho^m$ where $\rho$ is the total density of monomers, and
$m$ is the number of polymer chains involved in a contact,
hence $N_{inter}\propto \rho^2$ and $N_{intra}\propto \rho$. 
Every deviation from this ``trivial'' dependence thus also indicates
a deviation from mean field theory. Note that the ratio $N_{inter}/\rho$
in the bulk phases is the effective coordination number $z_{\rm eff}$
which we have used throughout this paper to calculate the Flory-Huggins
parameter $\chi$ from the interaction strength $\epsilon$ (eqn. (\ref{zeff})).
The profiles of $N_{inter}(z)/\rho(z)^2$ and $N_{intra}(z)/\rho(z)$
are shown for various $\epsilon$ in Fig. 15. According to the mean field 
assumption mentioned above, these quantities should be constant. In the 
simulations, they have a rather complex
structure. Generally, the chains rearrange in the vicinity of an interface
as to increase the number of intrachain contacts at the expense of the
interchain contacts. Right at the center of the interface, however,
the trend is opposite: The relative number of interchain contacts has
a maximum, and the number of intrachain contacts decreases.
This is presumably caused in part by the enrichment of chain ends at the 
interface. Furthermore, an additional entropic effect comes into
play very close to a sharp interface: Two monomers of the same chain which
are in contact are connected by a closed loop. If they are located in the 
immediate vicinity of a sharp interface, the loop can only extend into a 
half space which is entropically much less favorable than if the full space 
were available like further away from the interface. Thus, the number
of intrachain contacts is reduced at the interface.
We note that this fine structure of contact number profiles has not
been observed in previous studies of interfacial structures which
did not separate intrinsic profiles from capillary waves\cite{marcus1}.
Hence, this is the example of a case where intrinsic profile differ
{qualitatively} from their capillary wave broadened counterpart.
Otherwise, the capillary broadening does not affect the qualitative
shape of the profiles.

%
\section{Conclusions}

In this paper, we have presented extensive Monte Carlo simulations of
homopolymer interfaces, and analyzed them within the framework of a
theory\cite{capillary} which conceives the interface as a two dimensional 
surface embedded in space and decorated by intrinsic profiles which can be
obtained by mean field theory. We have shown that our results are
compatible with such a picture. The intrinsic profiles are in good
agreement with those obtained from the self-consistent field theory
on the length scale of a ``coarse graining length''  $B_0$.
The comparison between Monte Carlo simulation and self-consistent field 
calculations indicate that this length scale exhibits a dependence of the form 
$B_0 = 3.8 \: w_{\rm SCF} (1 - 3.1/\chi N)$.
In the long chain length limit this behavior is in qualitative agreement 
with the suggestion of Semenov\cite{sem1} $B = \pi w$, however, we find 
pronounced chain end corrections.
We have to note that our raw simulation data do not inevitably
lend themselves to such an interpretation. Nothing in the curves shown
in Fig. 10 indicates that there should be anything special about the
self-consistent field width $w_{\rm SCF}$, or about the coarse graining length 
$B$ for which $w_B=w_{\rm SCF}$. The capillary wave description seems to
be valid down to length scales much smaller than that, down to block
sizes of about $B \approx 4$. It is conceivable that the interface on
these length scales can still be described by intrinsic profiles which 
would then have nothing to do with the self-consistent field prediction.
On the other hand, the local interfacial structure will presumably not
decouple from the fluctuations of the interfacial position on length
scales smaller than the extension of the chains, and something analogous
to the convolution approximation is rather unlikely to be valid.
This point will need further consideration in the future.
So far, the main merit of our analysis is to have provided 
insight on the validity range of the self-consistent field
theory, and on the length scale on which it is applicable. 
A theory describing the local interfacial structure on length
scales much smaller than $w_{\rm SCF}$ has yet to be tested and established. 
For example, the P-RISM theory by Schweizer and Curro\cite{schweizer}
which puts much more emphasis on the local liquid structure,
is possibly a promising candidate. Such a theory will also hopefully
contribute to clarify the reason for the rather high
negative excess of total density in the interfacial region,
which we have discussed in section III.1.

\section*{Acknowledgments}

We have benefited from useful and stimulating discussions with 
T. Veitshans, W. Kob, T. Kerle, J. Klein.
The simulations were carried out on the computer facilities of the
ZDV, Mainz, the RHR, Kaiserslautern, and the Cray T3E at HLR, Stuttgart, and the HLRZ, J\"ulich.
Partial financial support by the Deutsche Forschungsgemeinschaft under 
grant Bi 314/3-4 and Bi 317, by the Materialwissenschaftliches Forschungszentrum Mainz 
(MWFZ), and by the Graduiertenkolleg on supramolecular systems in Mainz
is acknowledged.

\begin{figure}
\fig{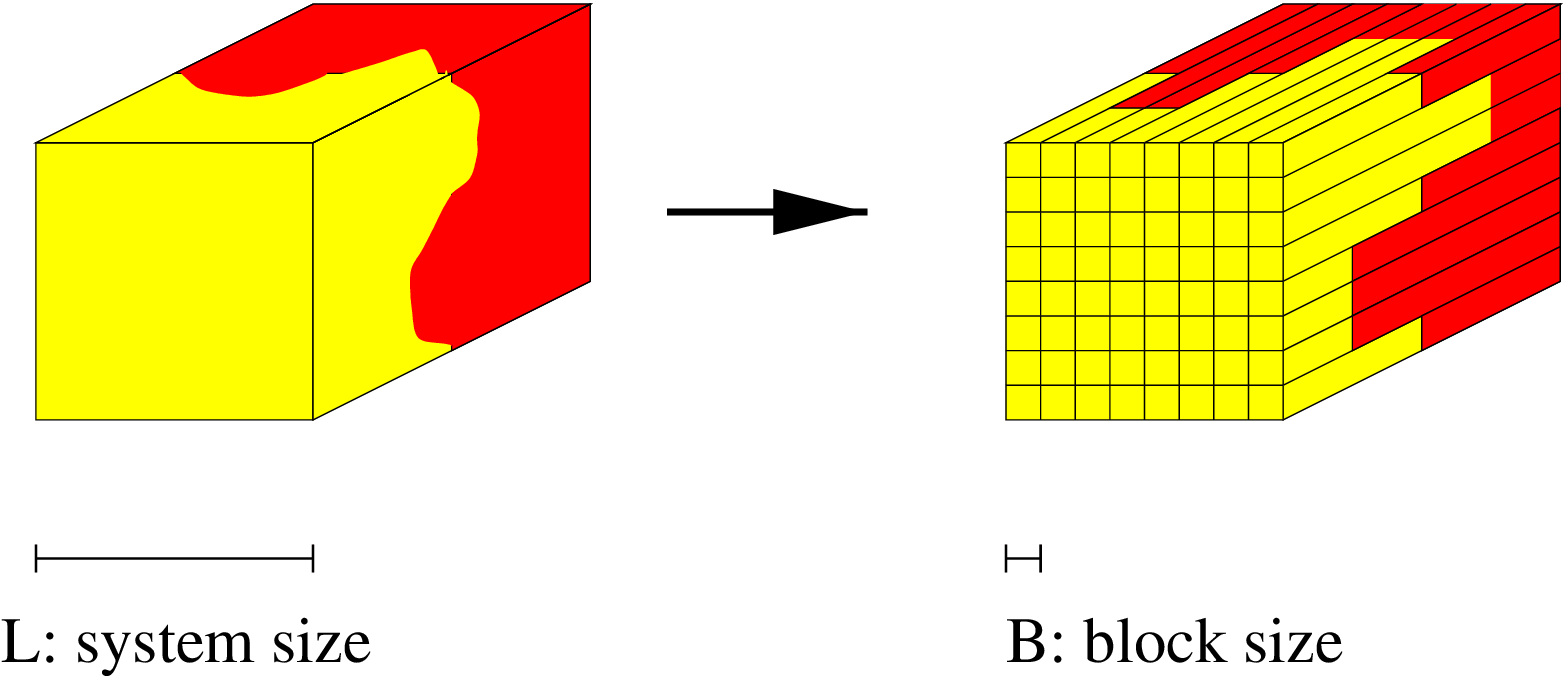}{150}{80}
\caption{\label{fig1}}
Block analysis, schematic picture.
\end{figure}

\begin{figure}
\fig{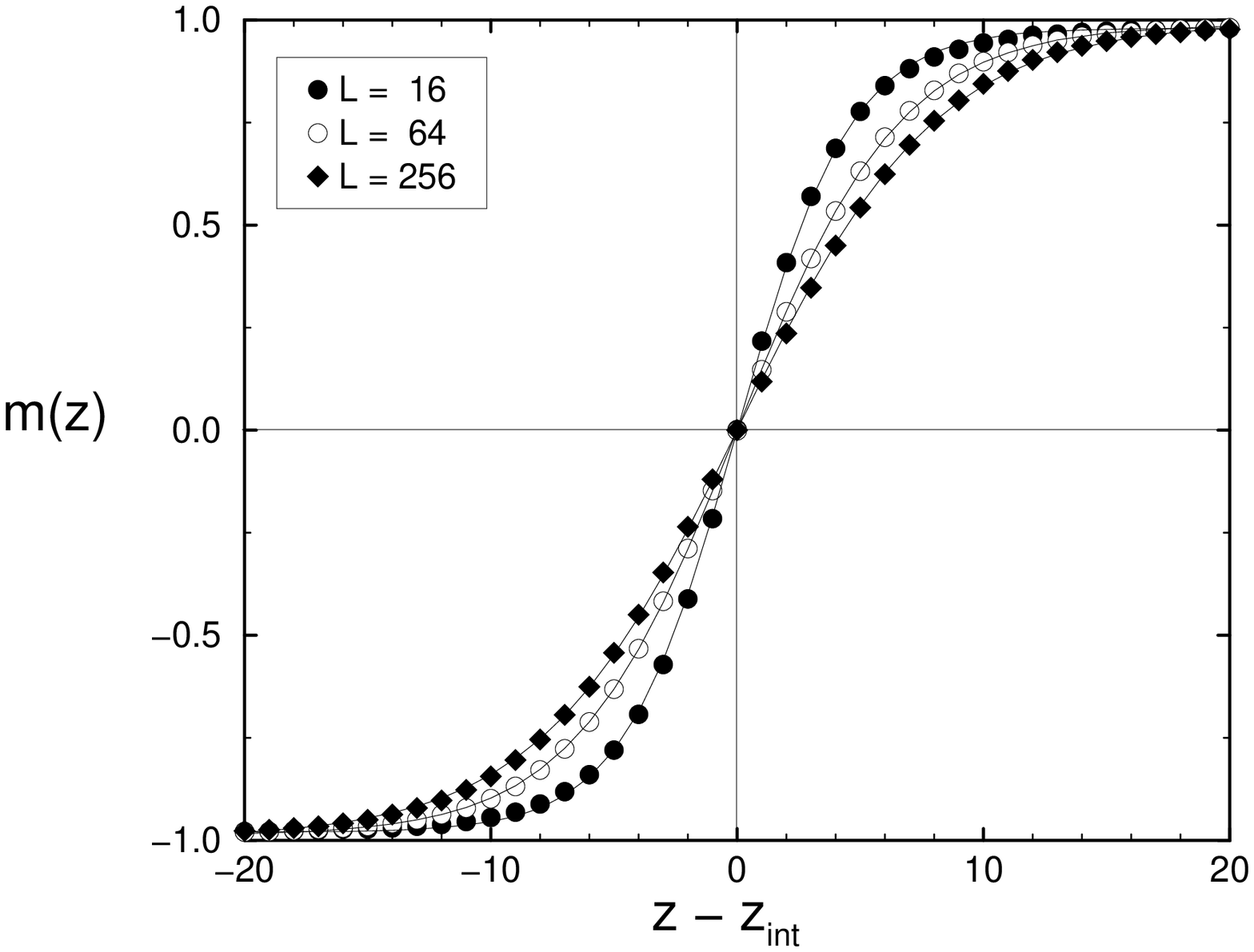}{100}{80}
\caption{\label{fig2}}
Apparent order parameter 
$m(z)=[\rho_A(z)-\rho_B(z)]/[\rho_A(z)+\rho_B(z)]$ vs. distance
from the interface position $z-z_{int}$ in units of the lattice constant, 
for different system sizes $L$.
Lines are fits to a tanh profile $m(z) = m_b \tanh[(z-z_{int})/w]$.
Parameters are $\epsilon=0.03$ and $N=32$
\end{figure}

\begin{figure}
\fig{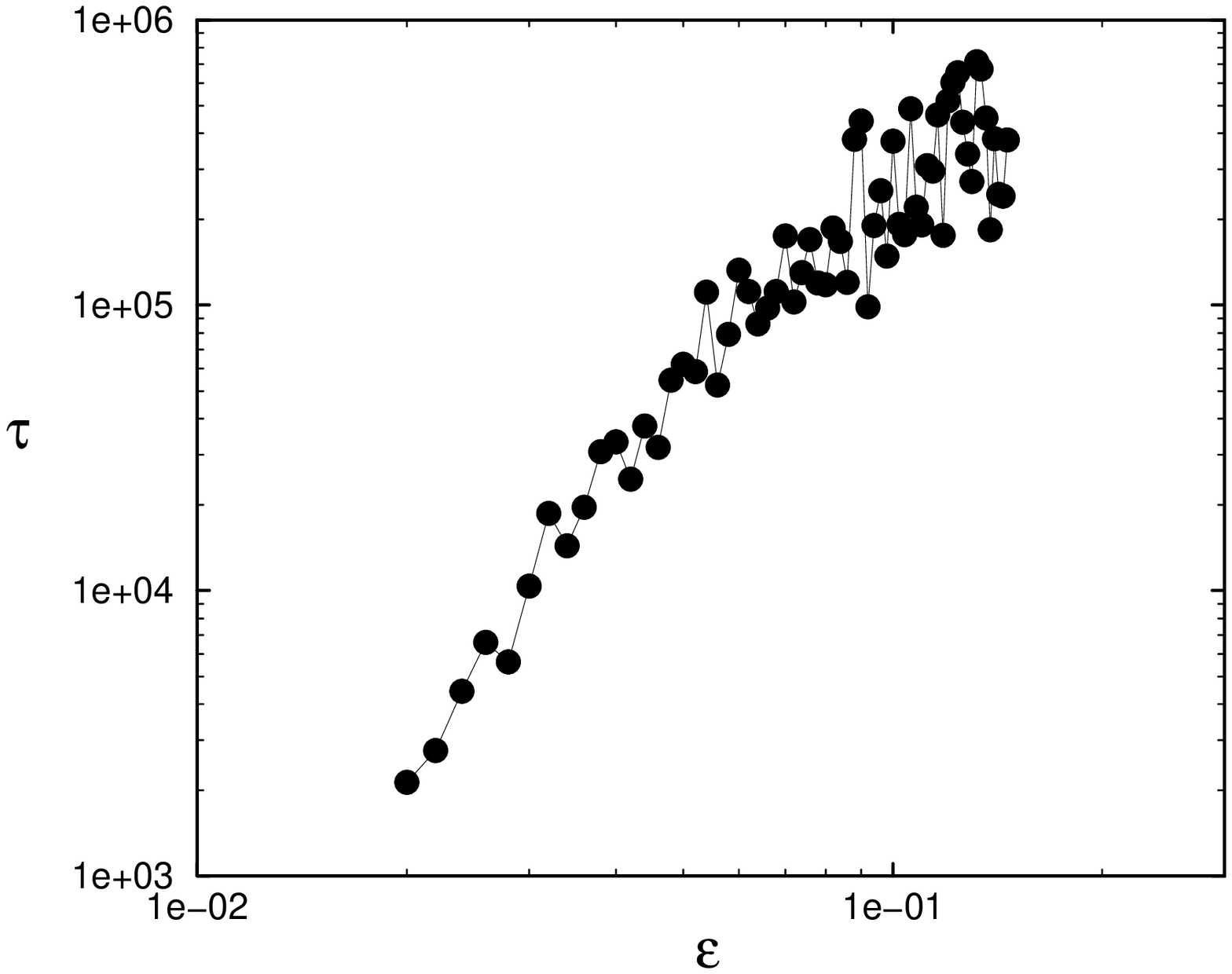}{100}{80}
\caption{\label{fig3}}
Autocorrelation time $\tau$ of the slowest capillary wave mode in units of 
Monte Carlo steps as a function of $\epsilon$ for chain length $N=32$. 
Four Monte Carlo steps correspond to one local hopping attempt per monomer, 
three slithering snake trials per chain and 
0.1 canonical particle exchange moves.
\end{figure}

\clearpage

\noindent
(a)\fig{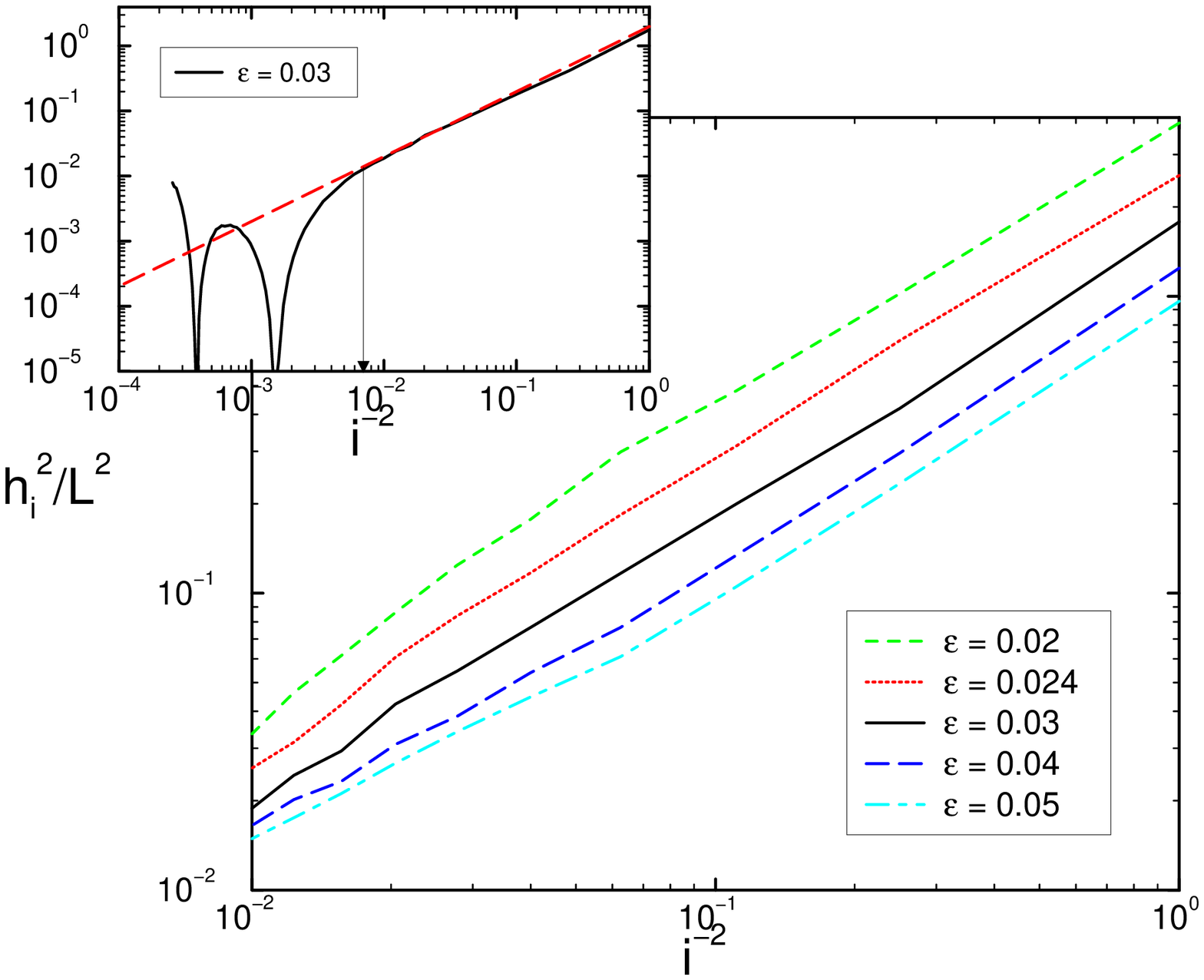}{100}{80}
(b)\fig{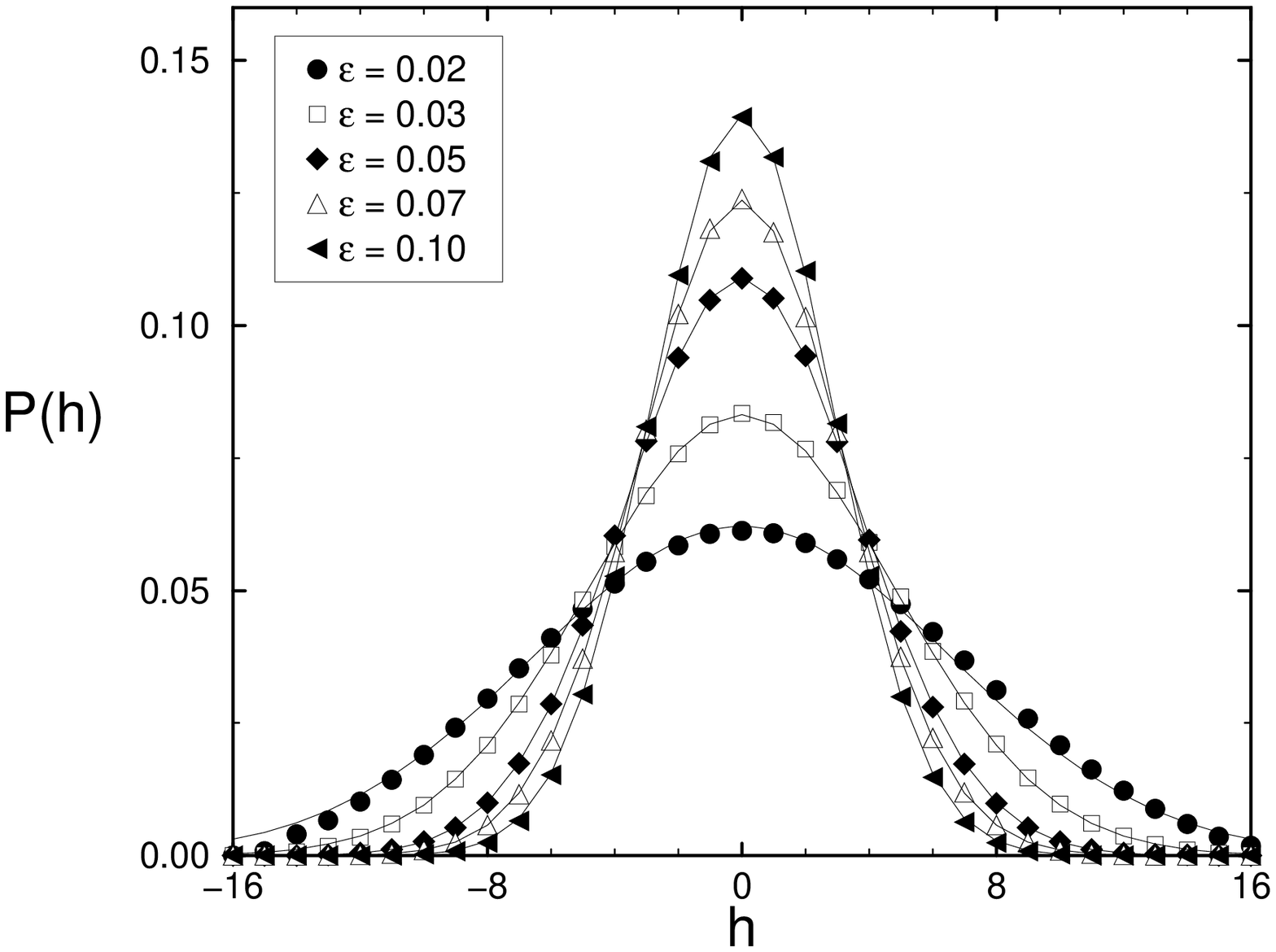}{100}{80}

\clearpage

\noindent
(c)\fig{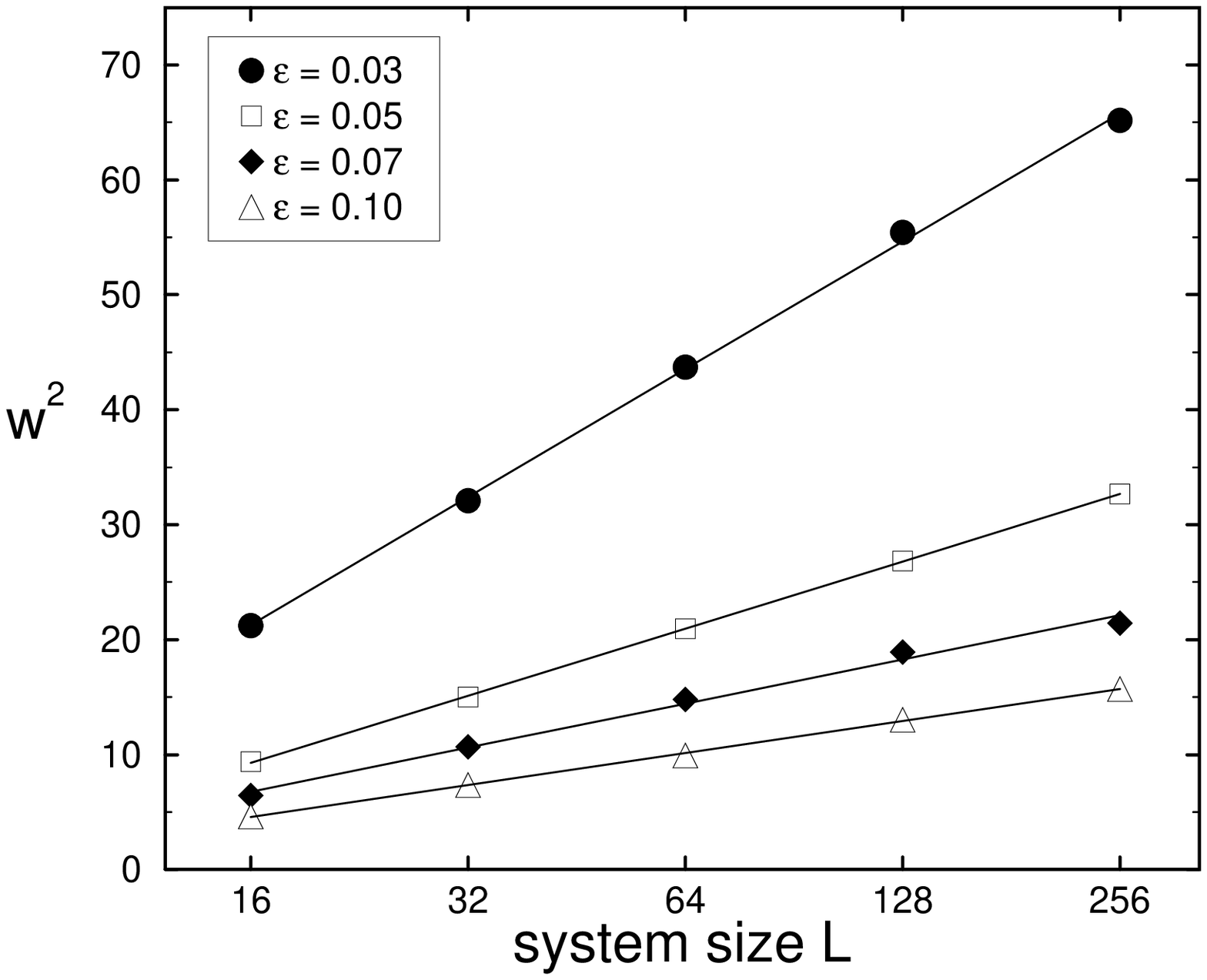}{100}{80}
(d)\fig{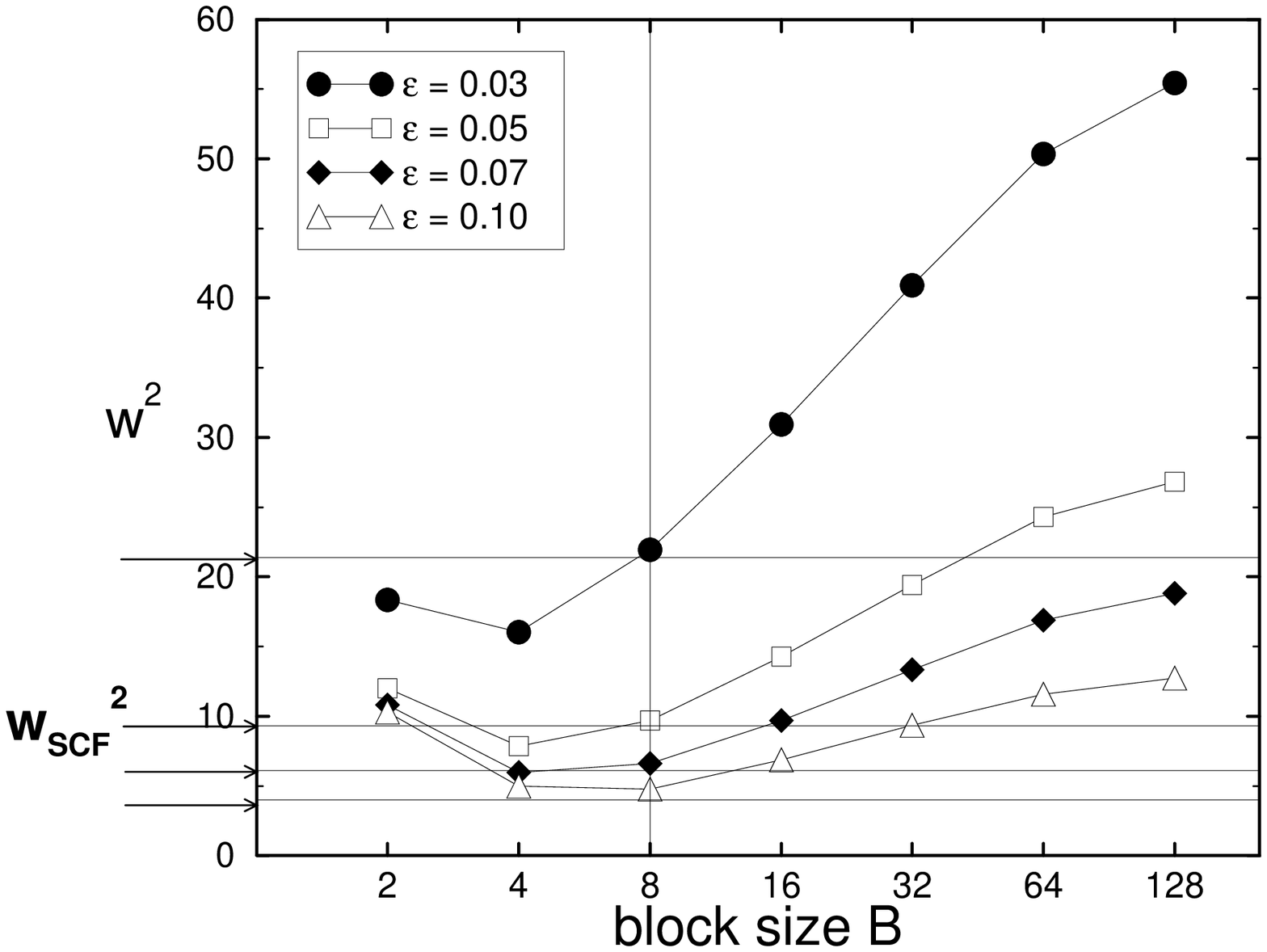}{100}{80}
\begin{figure}
\caption{\label{fig4}}

\clearpage

\centerline{FIG. 4}
Illustration of different strategies to extract the interfacial tension 
$\sigma$ from the capillary wave spectrum. Chain length is $N=32$,
film thickness $D=64$ (c) and $D=128$ ((a),(b), and (d)), 
system size $L=128$ except in (c), and
$\epsilon$ varies as indicated.  Lengths are given in units of the
lattice constant.  \\
(a) Double logarithmic plot of the amplitude $h_i{}^2$ of the $i$th Fourier 
mode in units of $1/L^2$ vs. $1/i^2$. Inset shows the whole
spectrum for $\epsilon = 0.03$. The capillary wave regime sets in at
$L/i = 10.7$ (arrow). The theoretical prediction in this regime is 
$\langle | h_i |^2 \rangle = [L/2 \pi i]^2 \sigma^{-1}$ (dashed line, inset).\\
(b) Distribution $P(h)$ of the local interface position at block size $B=8$.
Lines are fits to a Gaussian distribution $P(h) \propto \exp(-h^2/2 s^2)$.
The theoretical prediction is $s^2 = 1/(2 \pi \sigma) \ln (L/B)$.\\
(c) Apparent interfacial width $w^2$ as a function of system size $L$. 
Lines are fits to the theoretical prediction 
$w_L{}^2 = w_{intr.}{}^2 + 1/(4 \sigma) \ln(L/B)$.\\
(d) Apparent interfacial width $w^2$ as a function of block size $B$.
Arrows show the self-consistent field prediction $w_{\rm SCF}$. 
\end{figure}

\begin{figure}
\fig{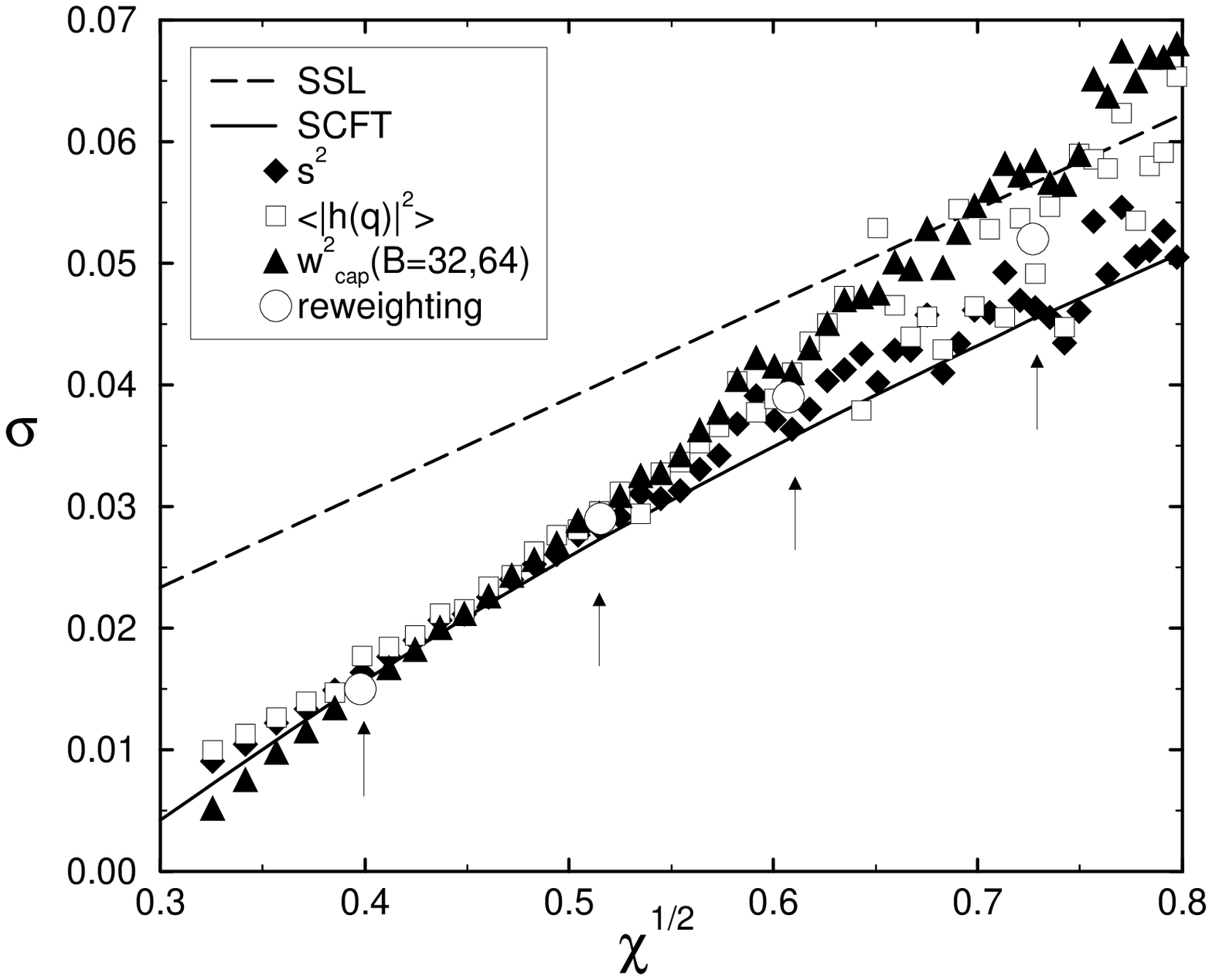}{100}{80}
\caption{\label{fig5}}
Interfacial tension $\sigma$ in units of $[$ lattice constant $]^{-2}$
as a function of Flory-Huggins parameter 
$\chi$, as obtained with the different methods illustrated in Fig. 4.
Also shown are independent values from M\"uller {\em et al}\cite{marcus1}, 
measured from bulk simulations with histogram reweighting techniques. 
The solid line shows the self-consistent field prediction and the dashed line
the strong segregation limit $\sigma_{SSL}=\rho b \sqrt{\chi/6}$.
\end{figure}

\begin{figure}
\fig{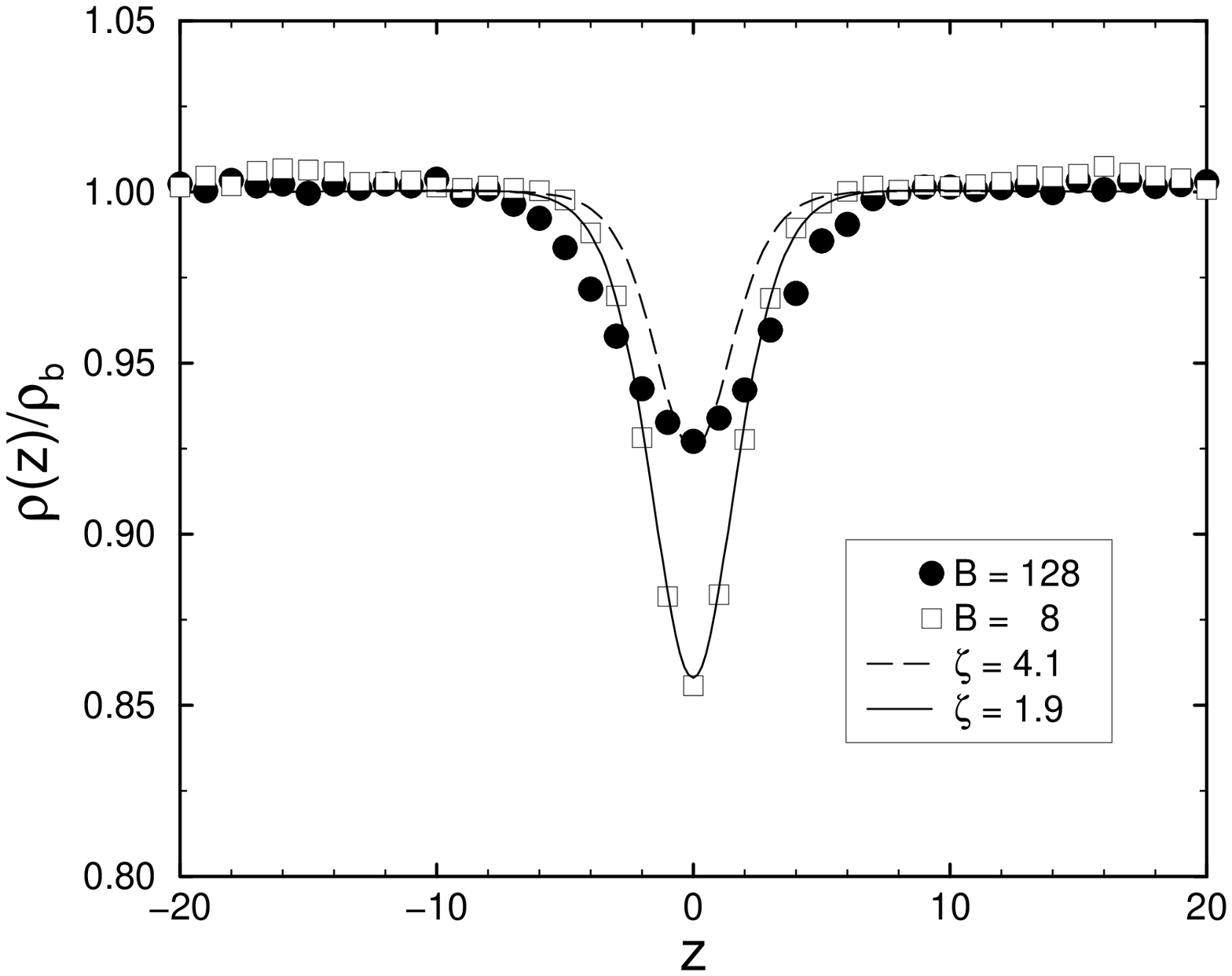}{100}{80}
\caption{\label{fig6}}
Total density profile $\rho(z)$ in units of $\rho_b$ vs. $z$ in units of
the lattice constant as
measured at block size $B=8$ (open squares) and $B=L=128$ (filled circles).
Lines show the self-consistent field prediction for the compressibility
parameter $\zeta=4.1$ (dashed) and $\zeta=1.9$ (solid).
Parameters are $N=32$ and $\epsilon=0.1$.
\end{figure}

\clearpage

\noindent
(a) \fig{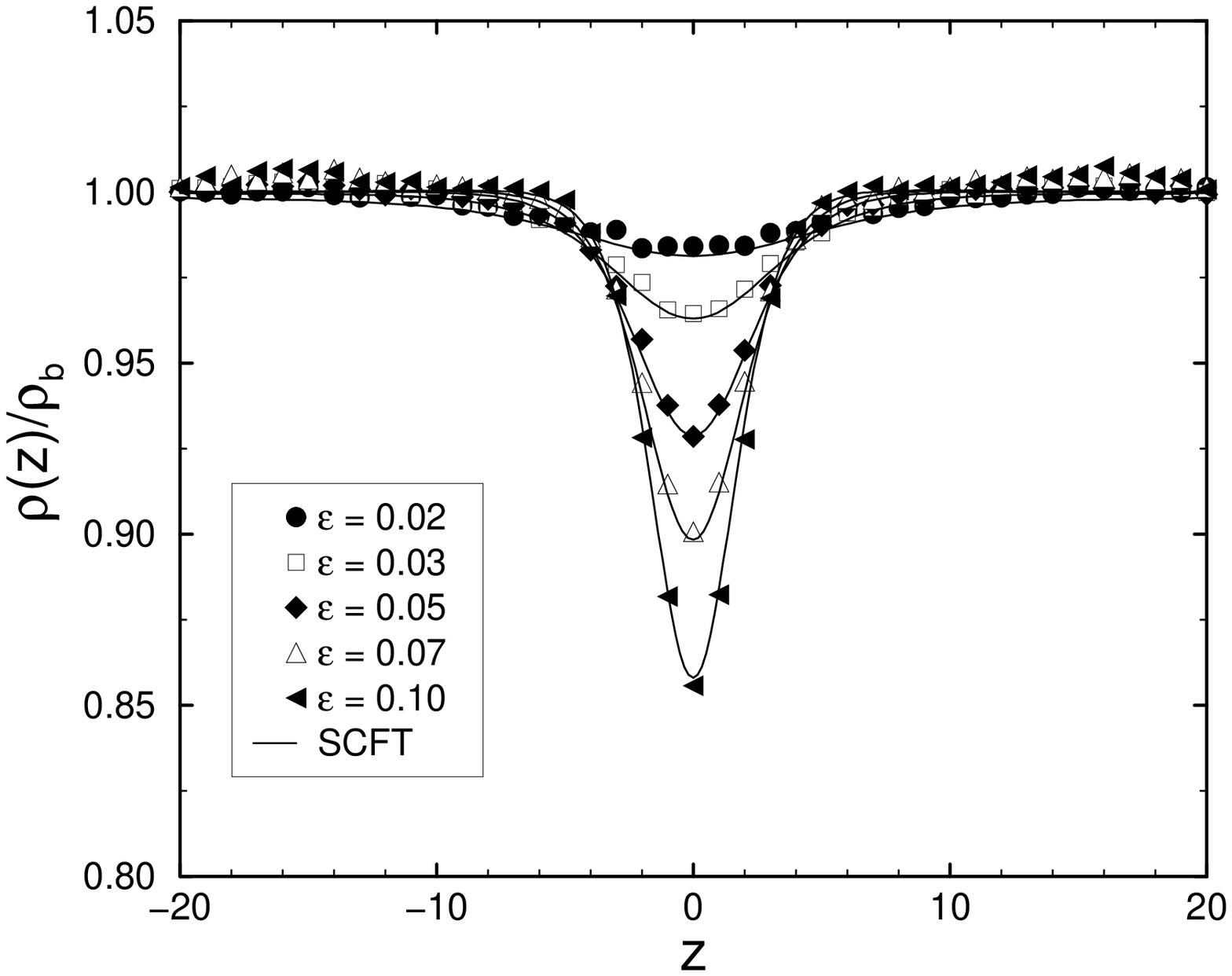}{100}{80}
(b) \fig{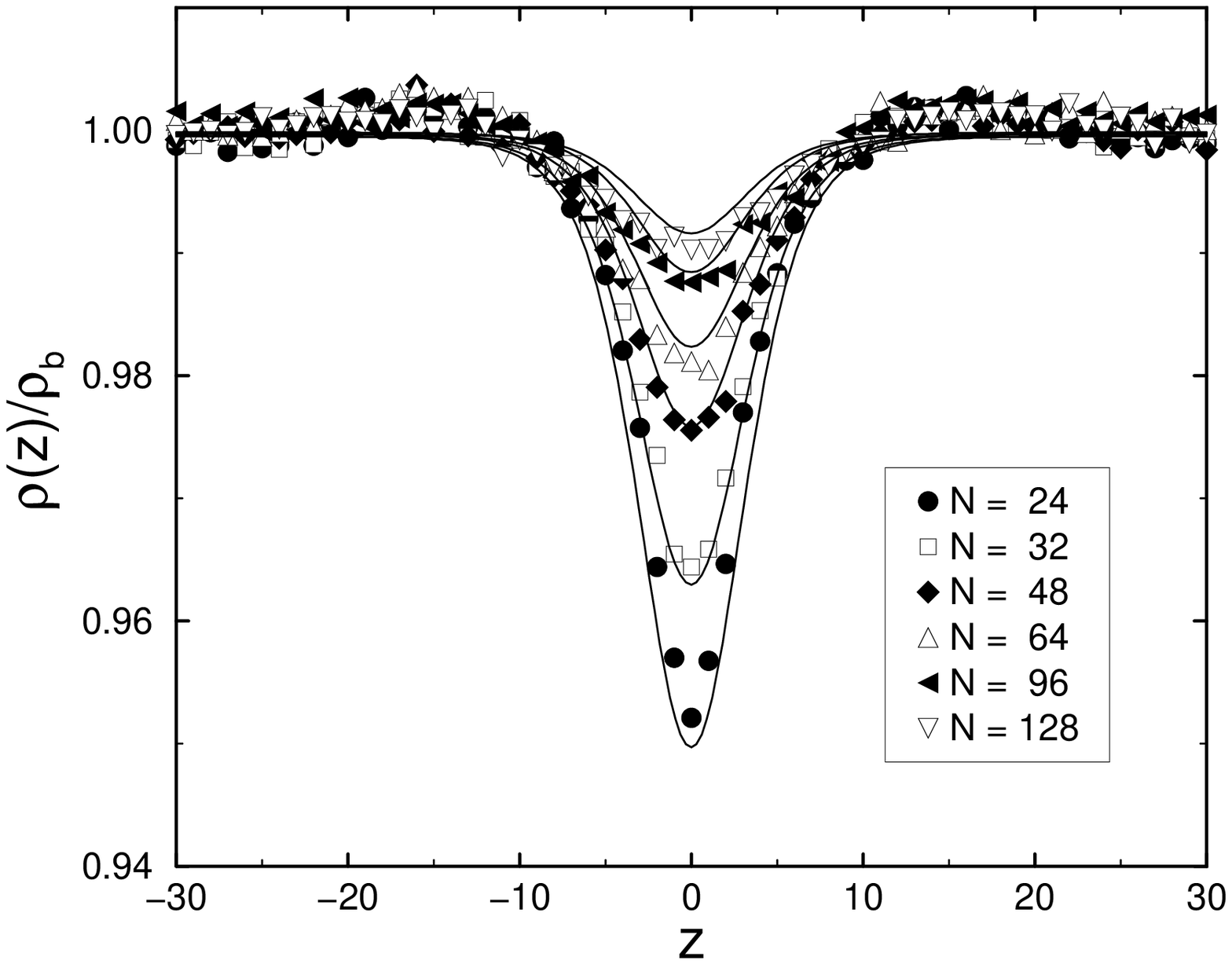}{100}{80}
\begin{figure}
\caption{\label{fig7}}
Total density profiles for block size $B=8$
(a) at chain length $N=32$ for different $\epsilon$
(b) at fixed $\epsilon N =0.96$ for different $N$.
Lines show the predictions of the self-consistent field theory using
$\zeta=1.9$. 
\end{figure}

\begin{figure}
\fig{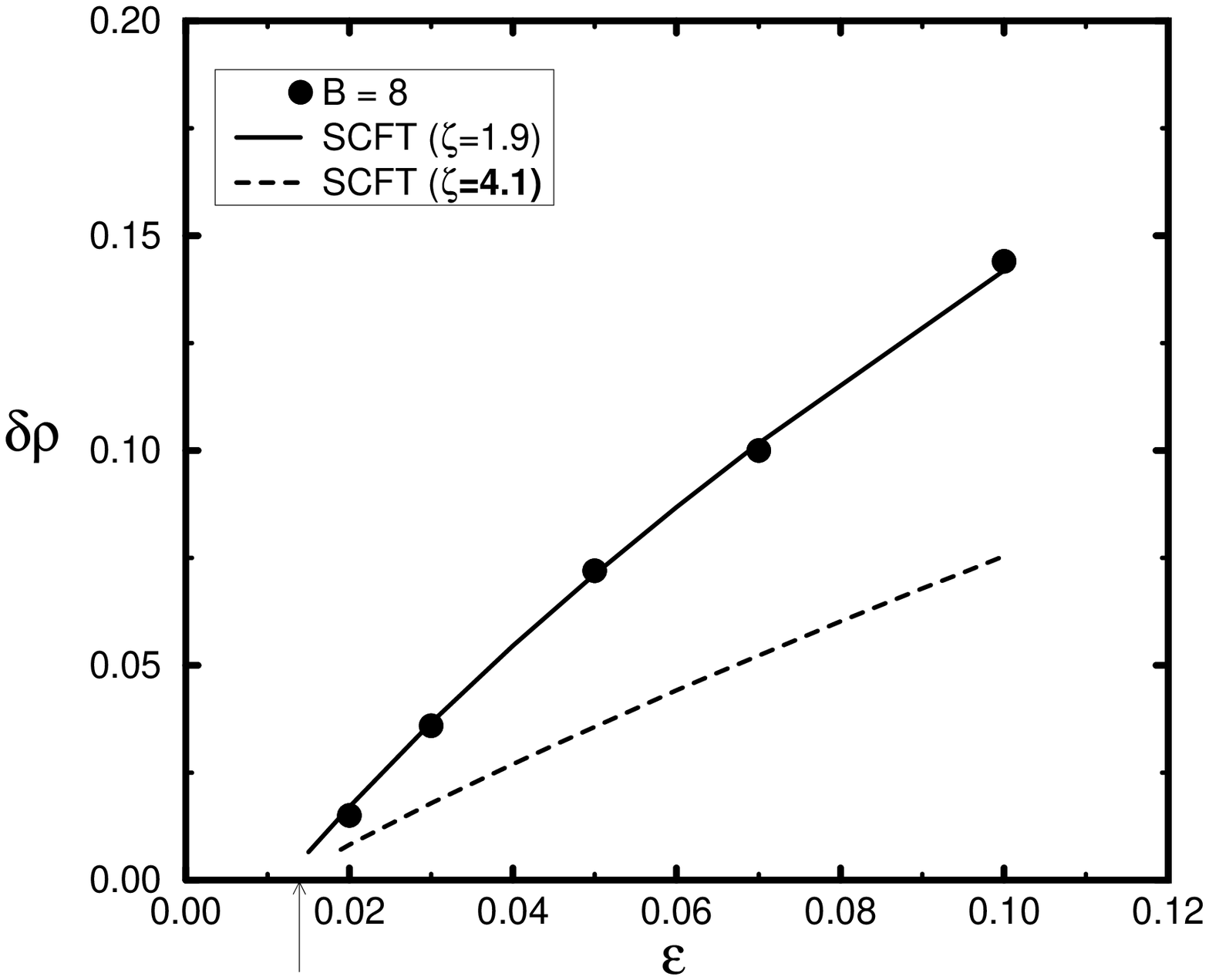}{100}{80}
\caption{\label{fig8}}
Depth of the density dip $\delta \rho = 1-\rho(0)/\rho_b$
vs. $\epsilon$ for the profiles obtained at block size $B=8$ (Fig. 7a),
compared to the self-consistent field prediction for $\zeta=1.9$
(solid line) and $\zeta=4.1$ (short dashed line). 
The arrow indicates the critical demixing value of $\epsilon$.
\end{figure}

\clearpage

\noindent
(a) \fig{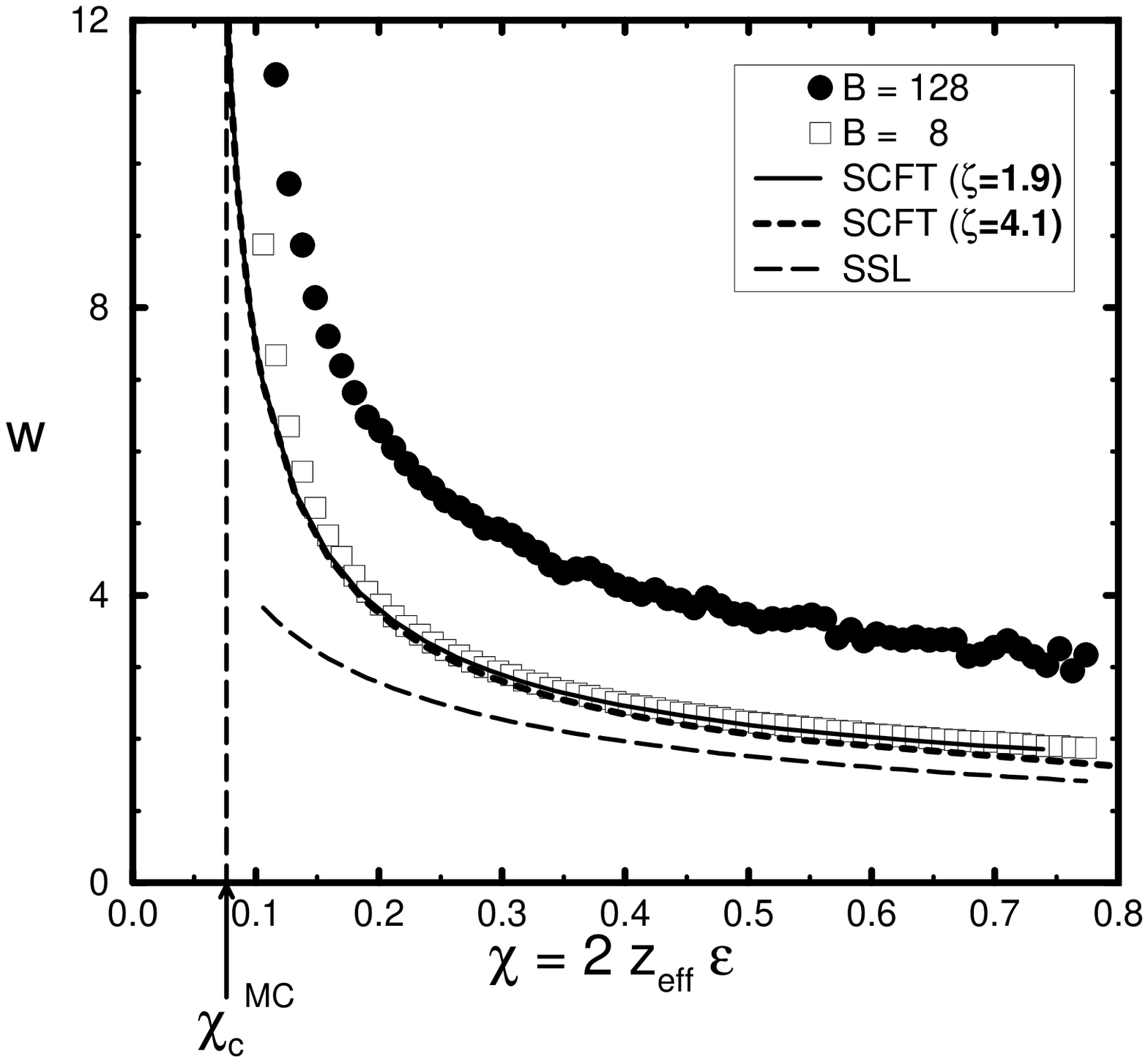}{100}{80}
(b) \fig{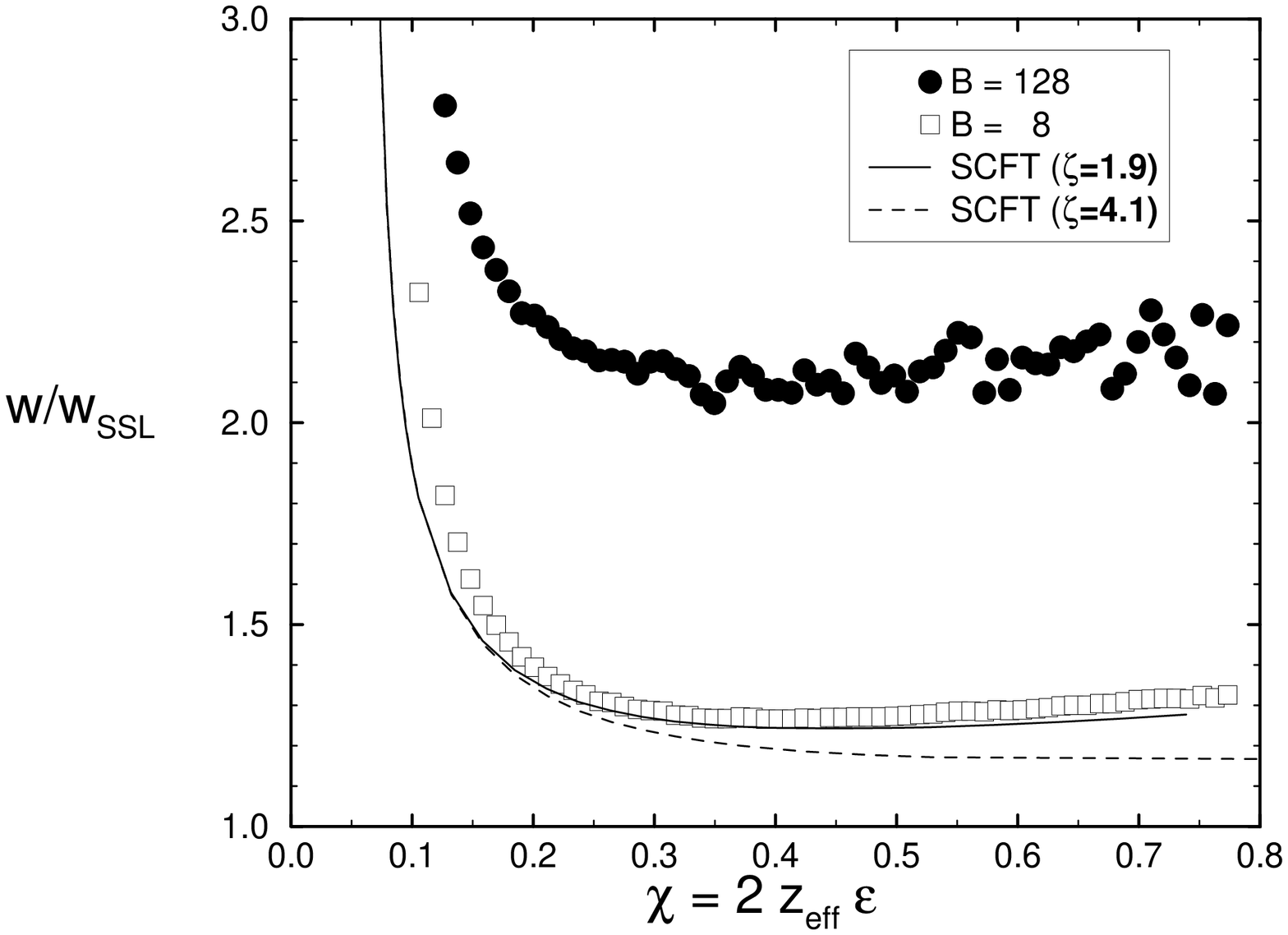}{100}{80}
\begin{figure}
\caption{\label{fig9}}
Interfacial width $w$ 
in absolute units (lattice constant) (a) and
in units of $w_{SSL}=b/\sqrt{6 \chi}$ (b)
at block size $B=8$ (open squares) and $B=L=128$ (filled circles)
vs. Flory-Huggins parameter $\chi$. Lines show the theoretical
prediction of the self-consistent field theory at $\zeta=1.9$ (solid),
$\zeta=4.1$ (short dashed), and the strong segregation limit
$w_{SSL}$ (long dashed). Arrow indicates critical demixing value $\chi_c$.
\end{figure}

\clearpage

\noindent
(a) \fig{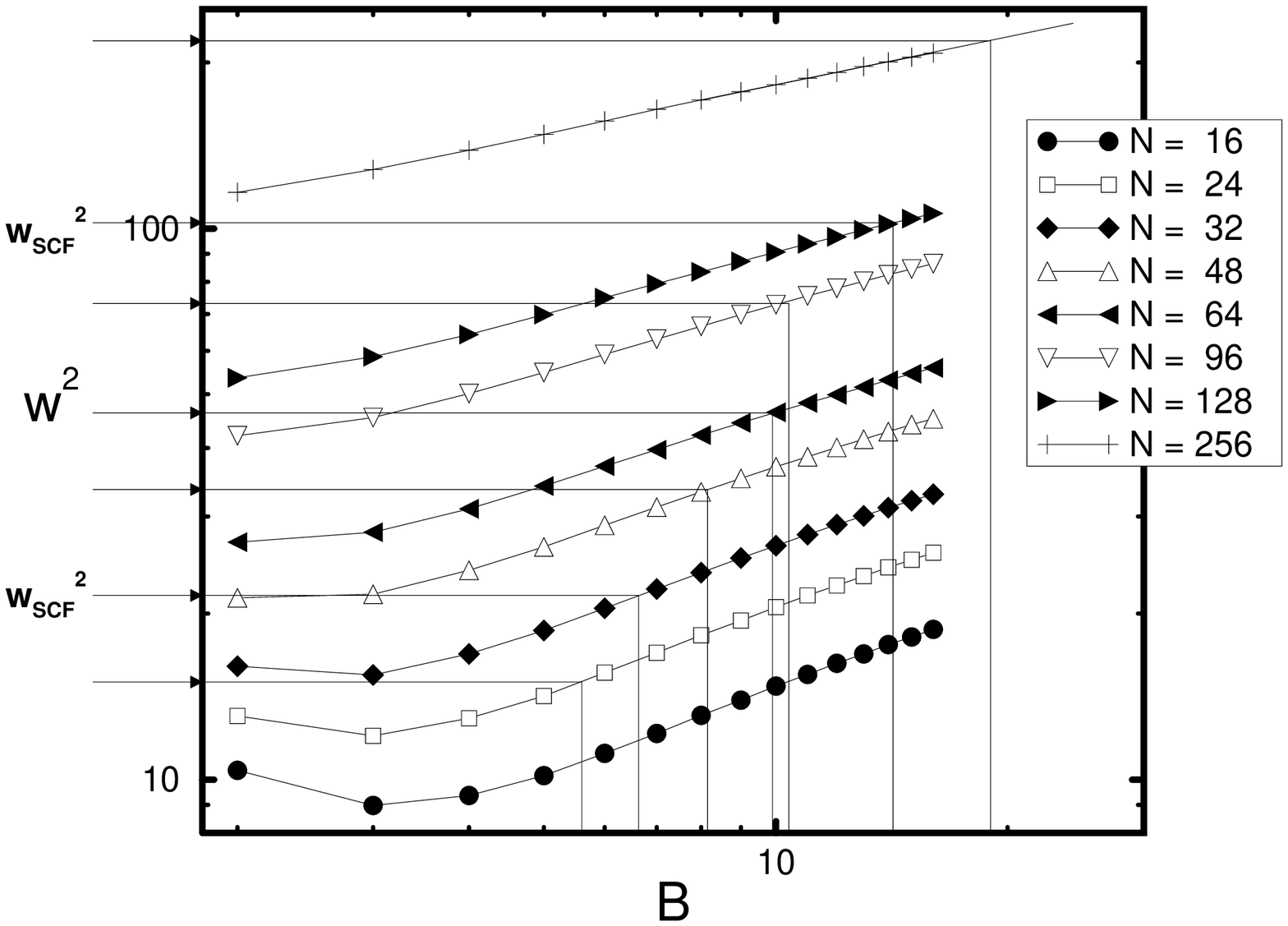}{100}{80}
(b) \fig{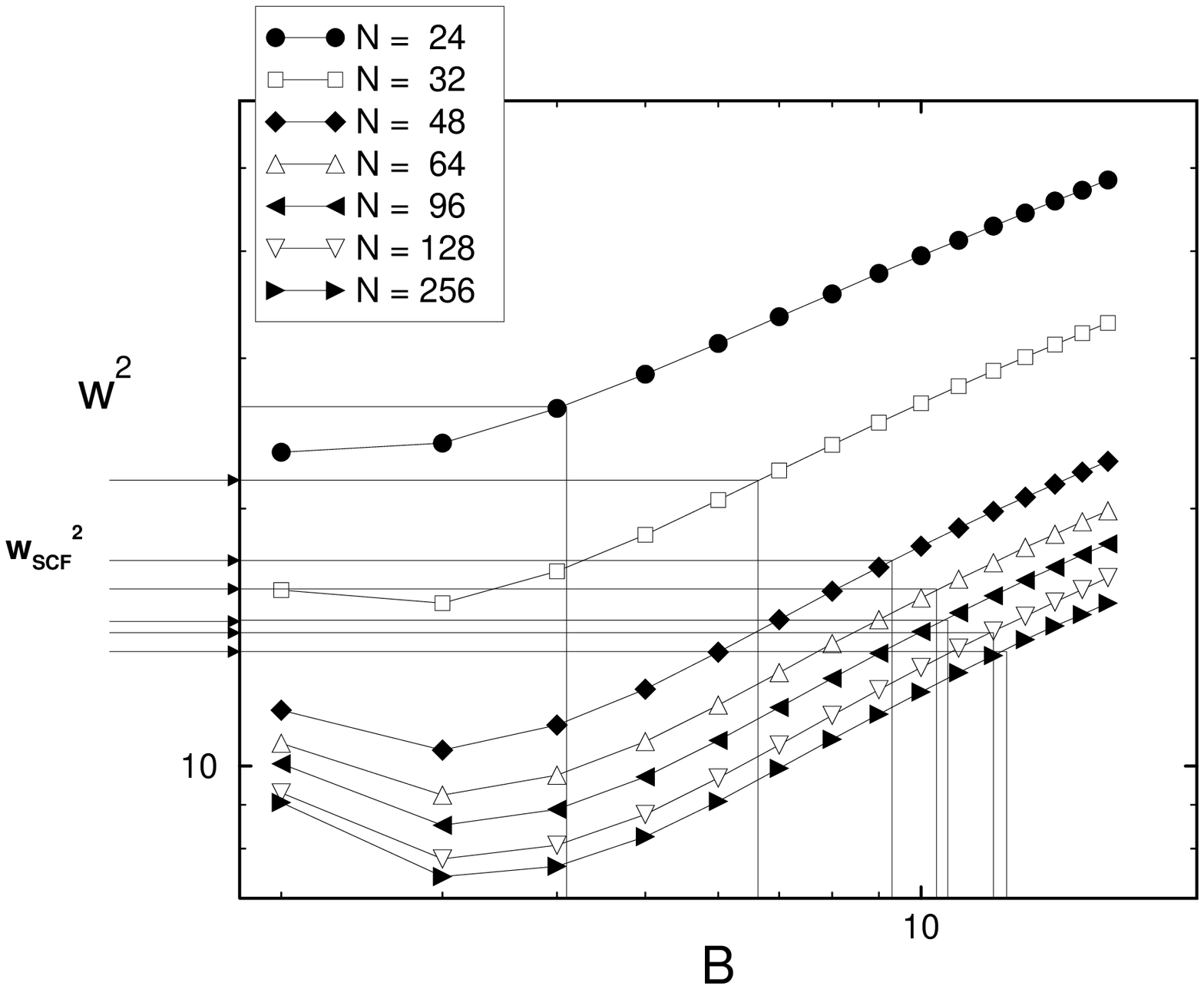}{100}{80}
\begin{figure}
\caption{\label{fig10}}
Squared apparent interfacial width $w^2$ as a function of block size $B$ in 
absolute units (lattice constants) for different chain lengths and
$\epsilon N=0.96=$ const. (a), $\epsilon = 0.03 =$ const. (b).
Arrows indicate the self-consistent field predictions. They are
used to read off the optimal block size $B$.
\end{figure}

\clearpage

\noindent
(a) \fig{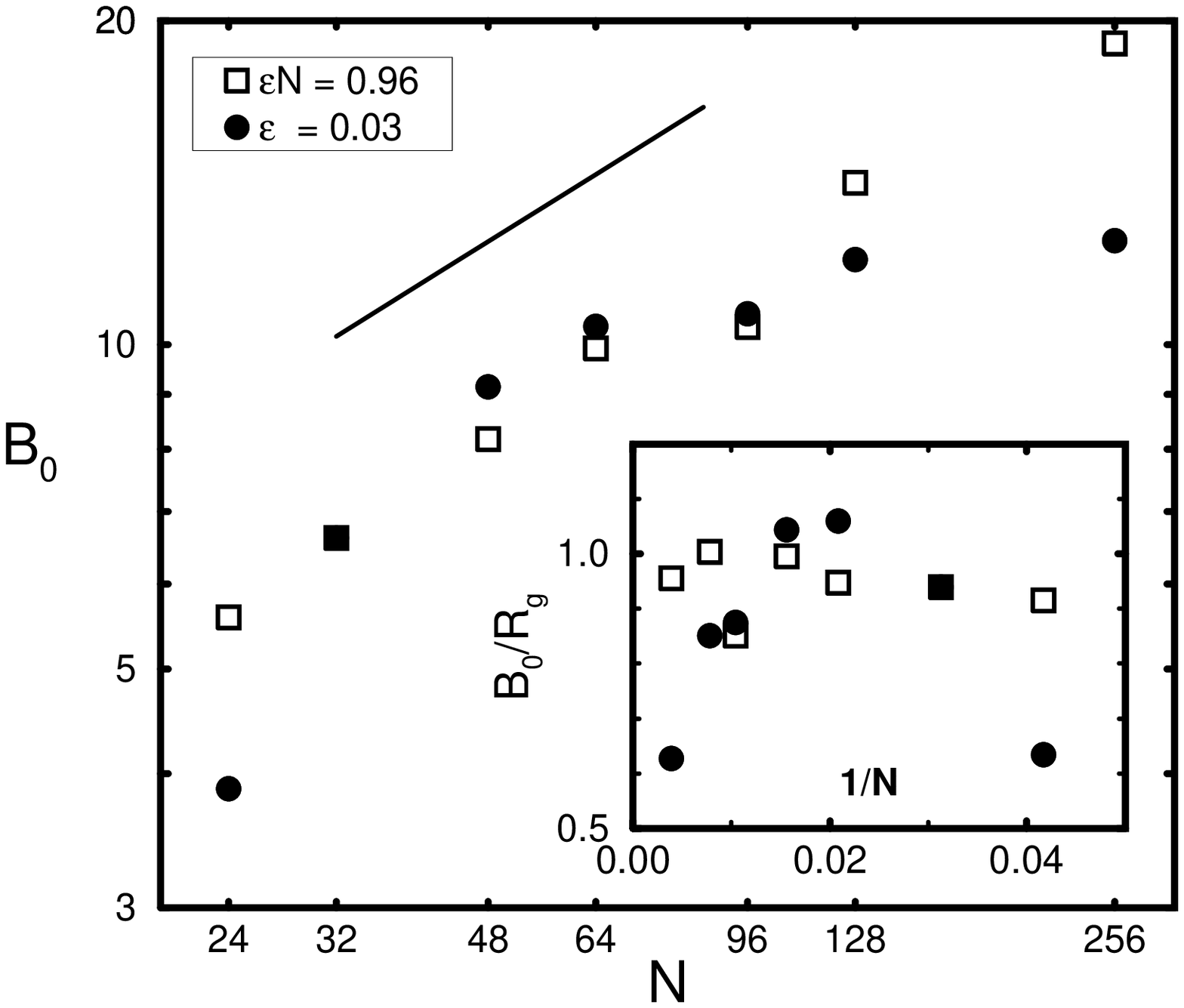}{100}{80}
(b) \fig{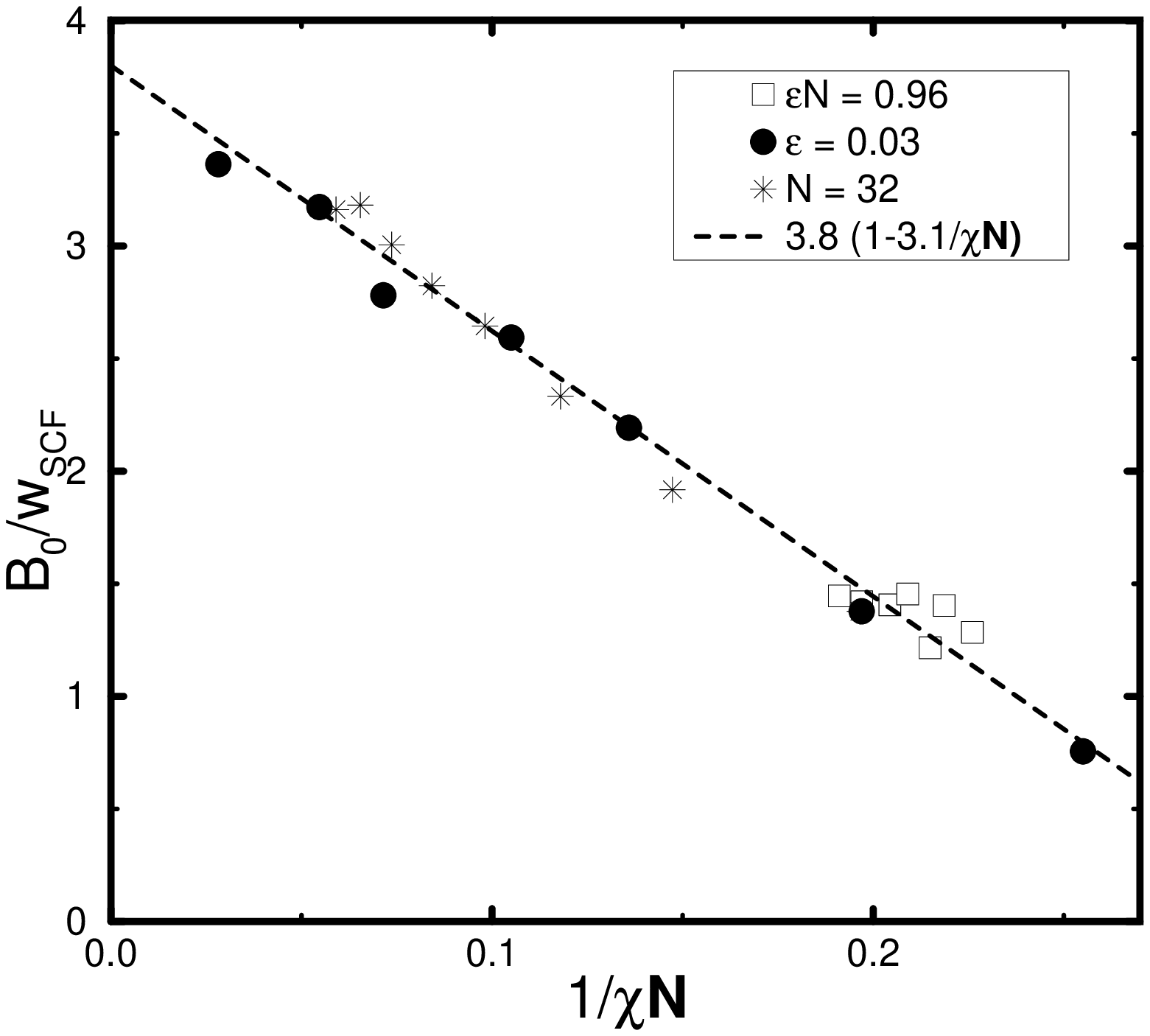}{100}{80}
\begin{figure}
\caption{\label{fig11}}
Optimal block size $B_0$ for different chain lengths $N$ and $\epsilon$
for $\epsilon N = 0.96 =$ const. (open squares),
$\epsilon = 0.03 =$ const. (filled circles), $N=32=$ const. (stars),
in absolute units as a function of chain length $N$ (a),
in units of $R_g$ as a function of $1/N$ (inset in (a)),
and in units of $w_{\rm SCF}$ as a function of $1/\chi N$ (b). 
Also shown in (a) is the slope 1/2 ($\sqrt{N}$), and in (b) a
fit of the data to $B = 3.8 w_{\rm SCF} (1-3.1/\chi N)$
\end{figure}

\begin{figure}
\fig{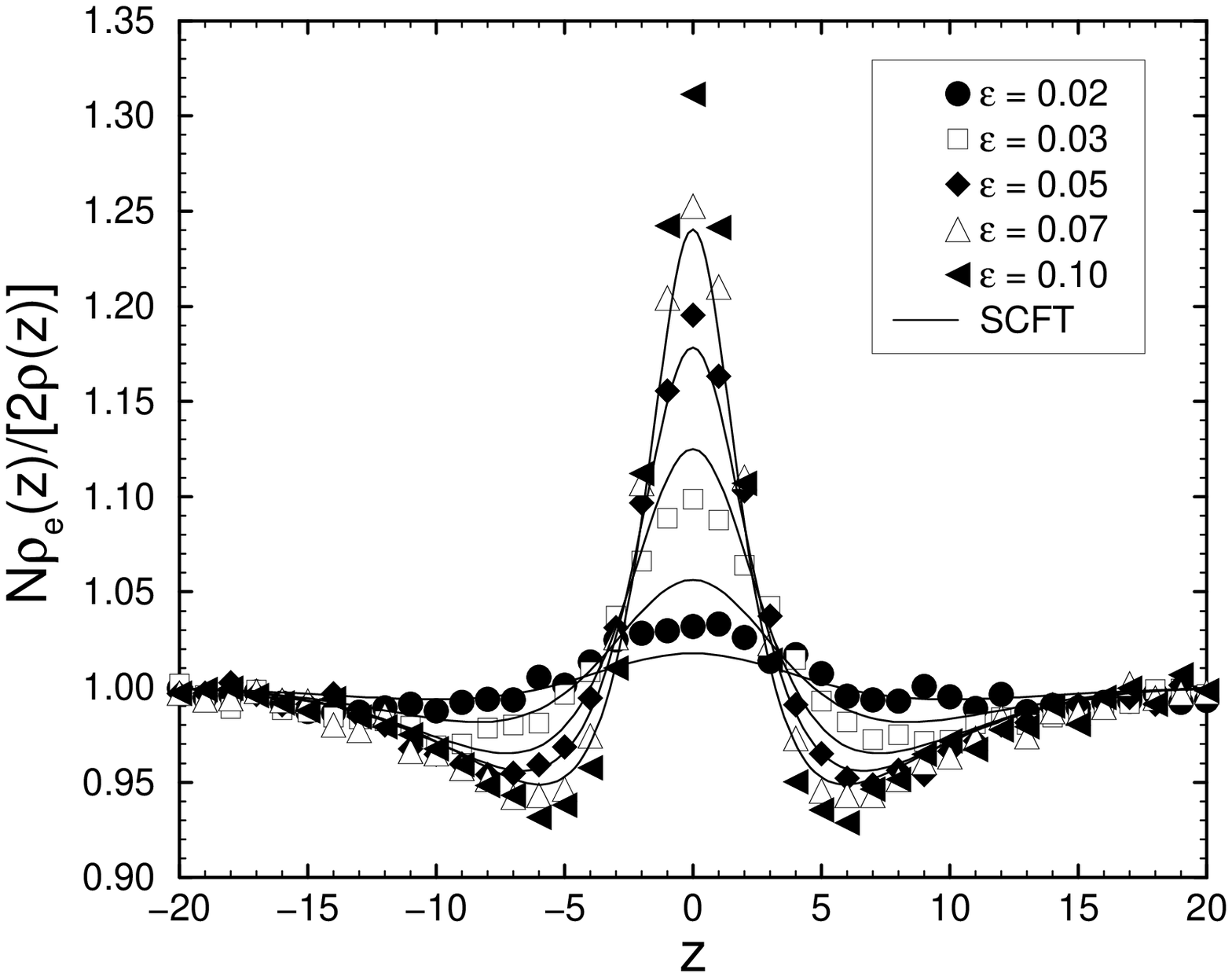}{100}{80}
\caption{\label{fig12}}
Normalized density of chain ends $\rho_e(z)$ vs. $z$ in units of the
lattice constant for chain length
$N=32$ and different $\epsilon$. Profiles were taken at block size $B=8$.
\end{figure}

\clearpage

\noindent
(a) \fig{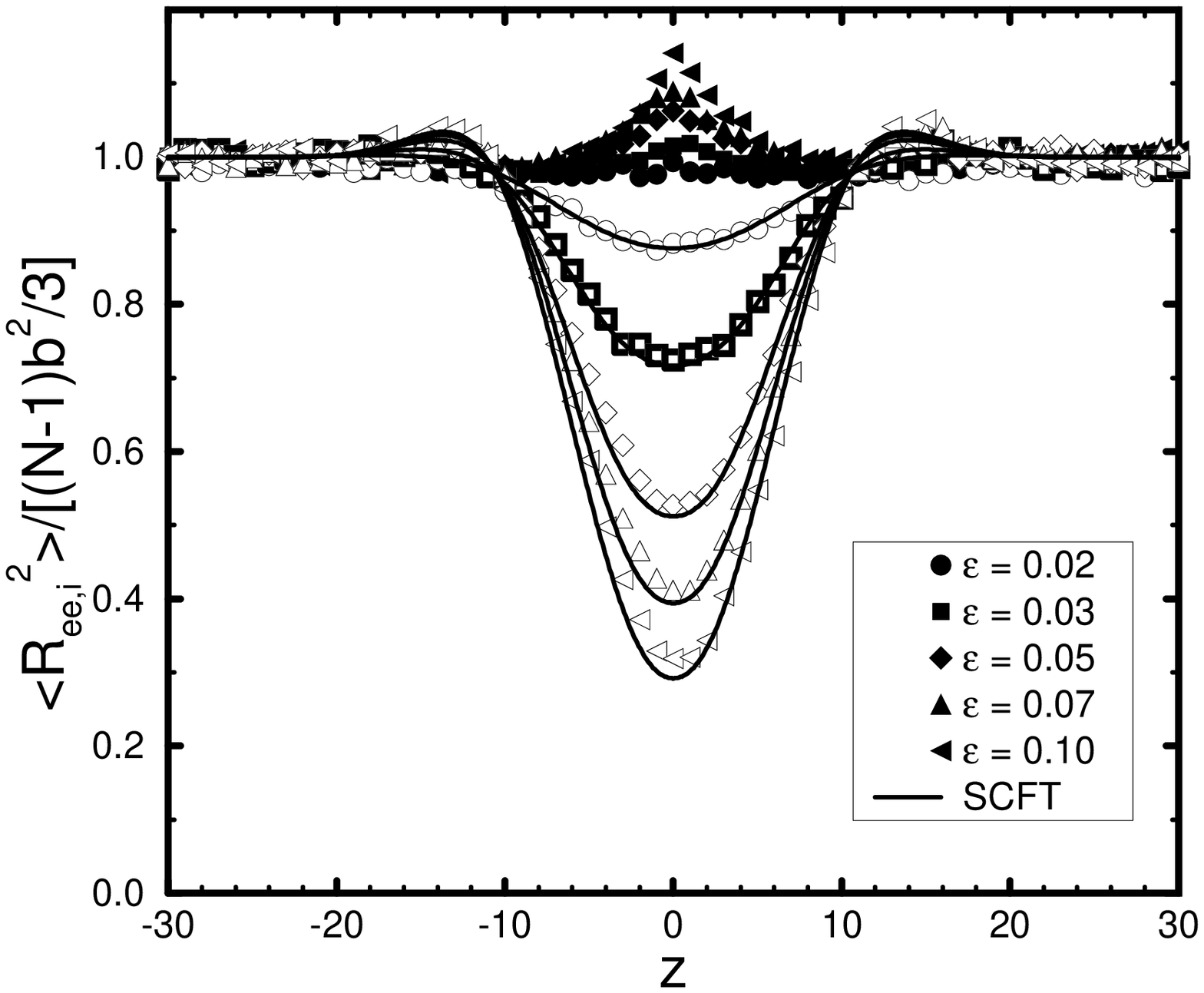}{100}{80}
(b) \fig{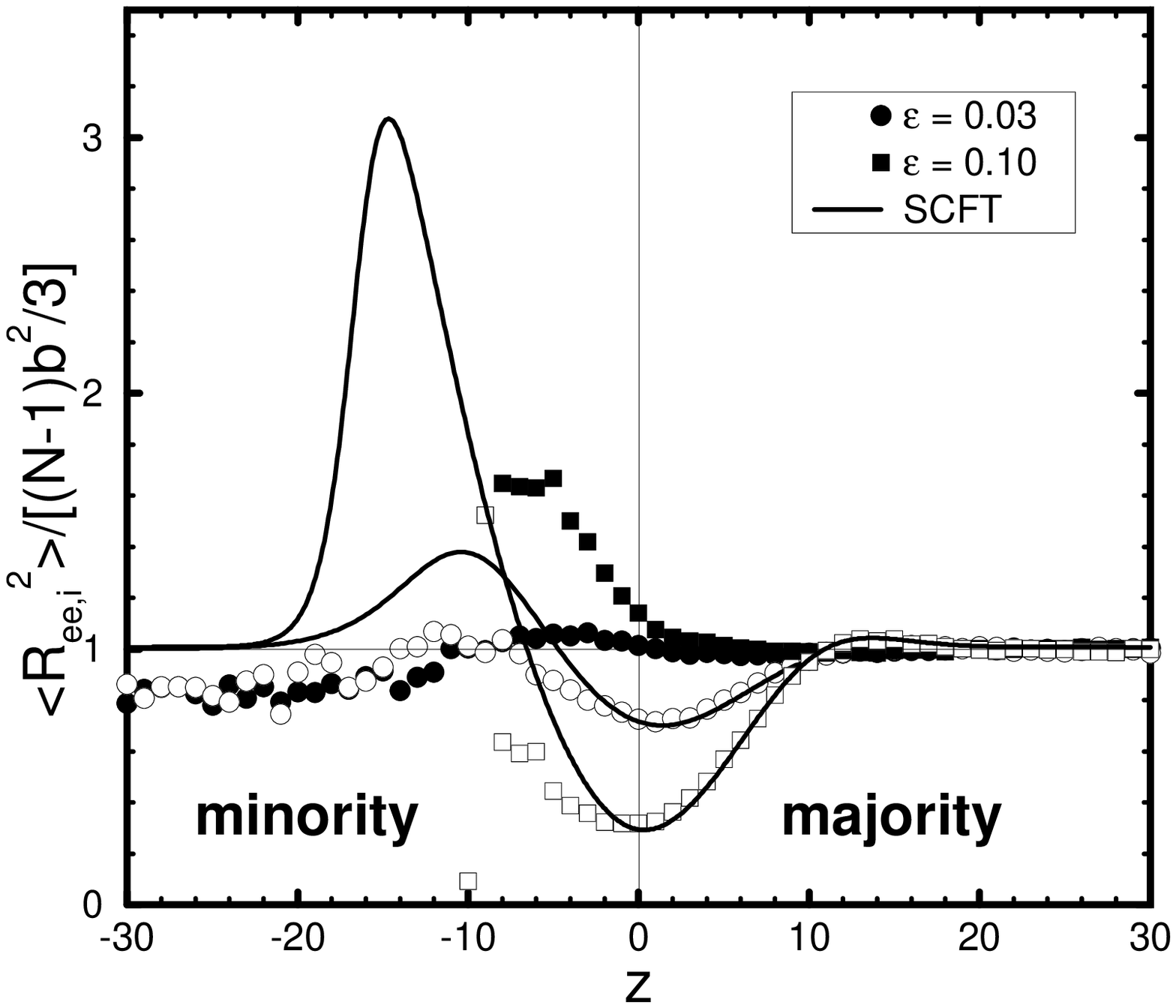}{100}{80}
\begin{figure}
\caption{\label{fig13}}
$x/y$ components (filled symbols) and $z$ components (open symbols) of
the squared end-to-end vector $\langle R_{ee,i}{}^2 \rangle$ in units
of the bulk value $b^2 (N-1)/3$ as a function of the distance $z$ of
the midpoint from the center of the interface (in units of the lattice
constant) for all chains,
(a) and for A chains only (b). Parameters are $N=32$, block size $B=8$ and 
$\epsilon$ as indicated. Lines show the predictions of the 
self-consistent field theory.
\end{figure}

\begin{figure}
\fig{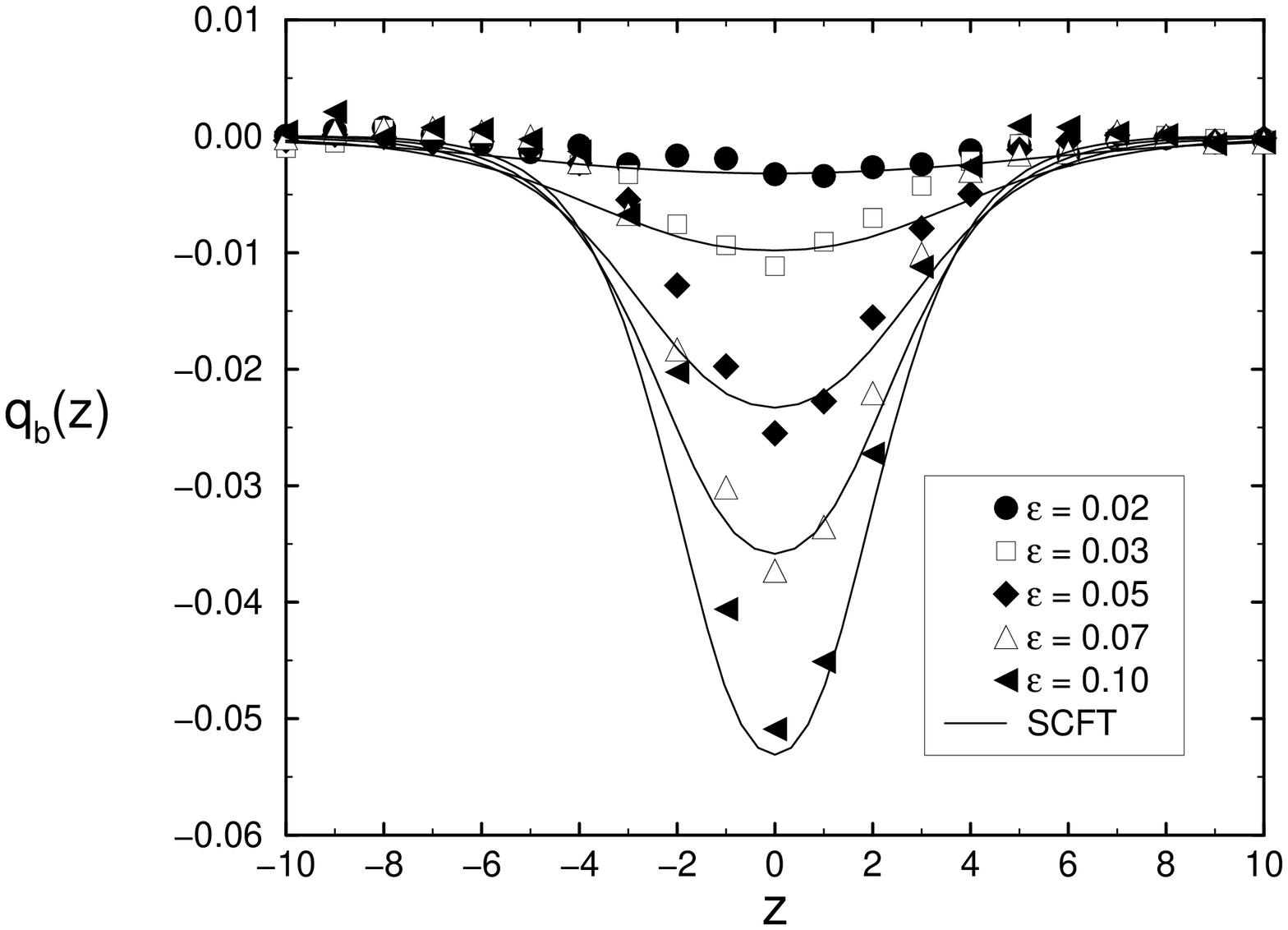}{100}{80}
\caption{\label{fig14}}
Bond orientational order parameter $q(z)$ vs. $z$ in units of the 
lattice constant for chain length $N=32$,
block size $B=8$, and $\epsilon$ as indicated. Lines show the predictions 
of the self-consistent field theory for a worm-like chain model with
chain stiffness $\eta=1.2$ (see text for explanation).
\end{figure}

\clearpage

\noindent
(a) \fig{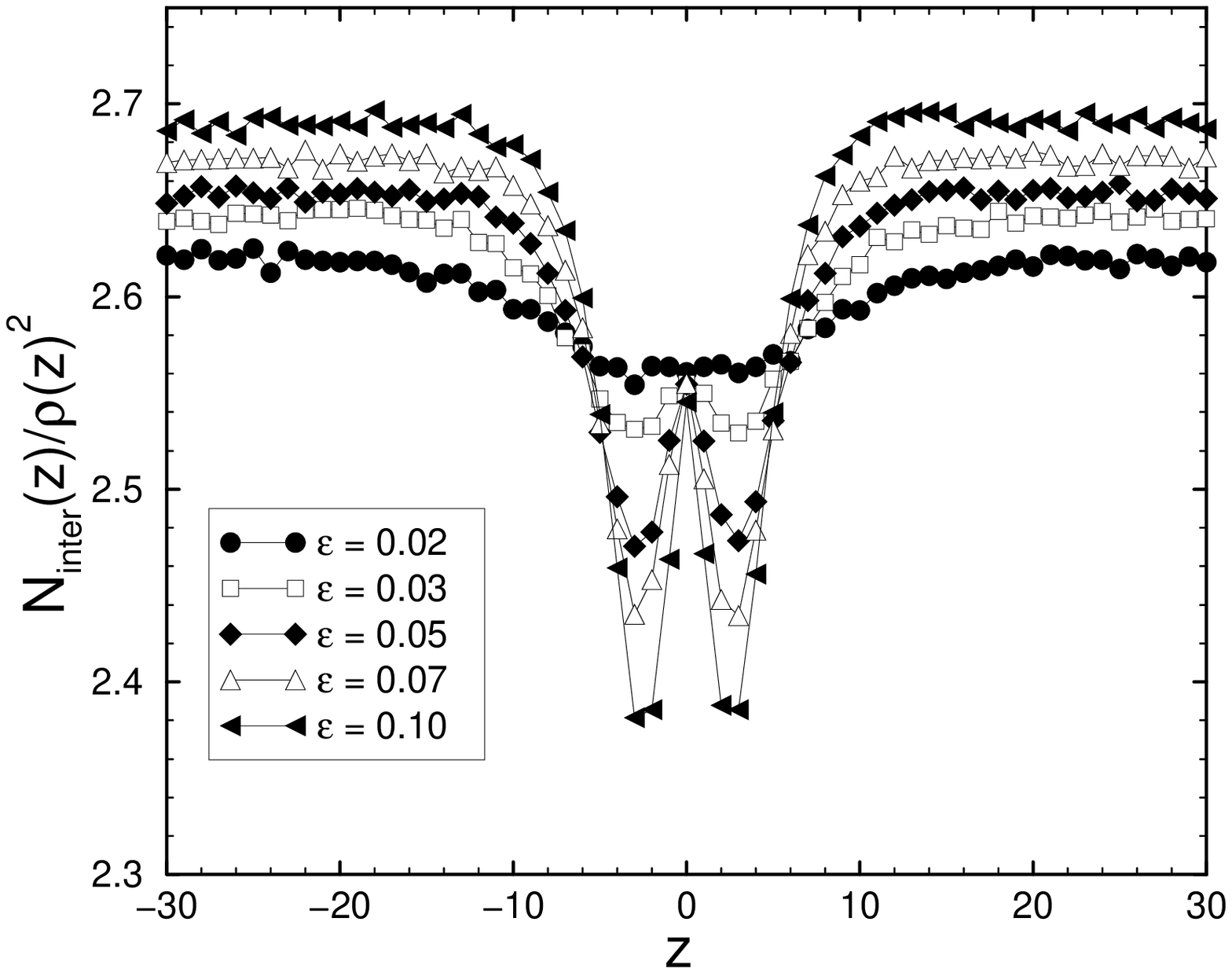}{100}{80}
(b) \fig{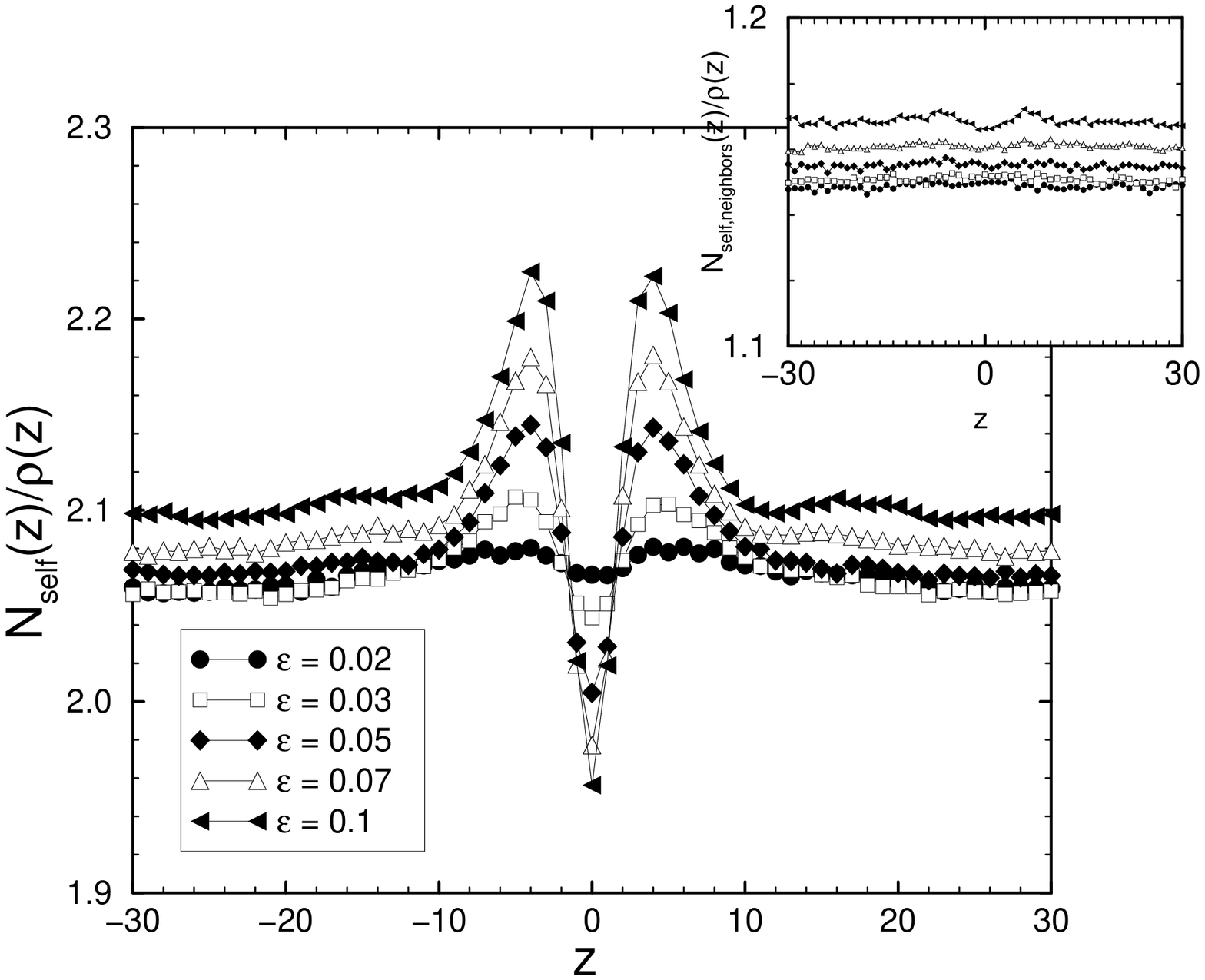}{100}{80}
\begin{figure}
\caption{\label{fig15}}
Profiles of 
(a) normalized number of interchain contacts $N_{inter}(z)/\rho(z)^2$ vs. $z$ 
in units of the lattice constant
and of (b) normalized number of self contacts $N_{self}(z)/\rho(z)$ vs. $z$ .
for different $\epsilon$. Inset in (b) shows the number of contacts
with direct neighbors along the chain.  Parameters are $N=32$ and $B=8$.
\end{figure}



\begin{thebibliography}{99}
\bibitem{blends} 
D.R. Paul and S. Newman,
{\it Polymer Blends},
Academic Press, New York (1978);
K. \v{S}olc (edt.),
{\it Polymer Compatibility and Incompatibility},
Harwood Academic Publishers, Chur (1980);
D.S. Walsh, J.S. Higgins, and A. Maconnachie,
{\it Polymer Blends and Mixtures},
Martinus Nijhoff Publishers, Dordrecht (1985);
E.L. Thomas (edt.),
{\it Materials Science and Technology, Vol. 12: Structures and Properties of Polymers},
VCH, Weinheim (1993).
\bibitem{blends2}
K. Binder,
{\it Adv. Polymer Sci.}
{\bf 112}, 181 (1994).
\bibitem{ginzburg}
V.L. Ginzburg,
{\it Sov. Phys. Solid state}
{\bf 1}, 1824 (1960);
K. Binder,
{\it Phys. Rev.}
{\bf A 29}, 341 (1984).
\bibitem{capillary}
F. P. Buff, R. A. Lovett, F. H. Stillinger,
{\it Phys. Rev. Lett.}
{\bf 15}, 621 (1965);
J. D. Weeks,
{\it J. Chem. Phys.}
{\bf 67}, 3106 (1977);
D. Bedeaux, J. D. Weeks,
{\it J. Chem. Phys.}
{\bf 82}, 972 (1985).
\bibitem{sem1}
A.N. Semenov,
{\it Macromolecules}
{\bf 26}, 6617 (1993);
ibid
{\bf 27}, 2732 (1994).
\bibitem{shull}
K.R. Shull, E.J. Kramer, 
{\it Macromolecules} 
{\bf 23}, 9769 (1990).
\bibitem{kerle}
T. Kerle, J. Klein, and K. Binder,
{\it Phys. Rev. Lett.}
{\bf 77}, 1318 (1996);
T. Kerle, J. Klein, and K. Binder,
submitted to 
{\it Eur. Phys. J.}.
\bibitem{sferraza}
M. Sferrazza, C. Xiao, R.A.L. Jones, D.G. Bucknall, J. Webster, and J. Penfold,
{\it Phys. Rev. Lett.}
{\bf 78}, 3693 (1997).
\bibitem{helfand}
E. Helfand and Y. Tagami,
{\it J. Polym. Sci., Polym. Lett.}
{\bf 9}, 741 (1971);
{\it J. Chem. Phys.}
{\bf 56}, 3592 (1971);
ibid
{\bf 57}, 1812 (1972);
E. Helfand.
{\it J. Chem. Phys.}
{\bf 62}, 999 (1975).
\bibitem{flory}
P.J. Flory,
{\it Principles of Polymer Chemistry},
Cornell University Press, Ithaca (1953);
{\it J. Chem. Phys.}
{\bf 10}, 51 (1942);
M.L. Huggins,
{\it J. Phys. Chem.}
{\bf 46}, 151 (1942).
\bibitem{freed}
J. Dudowicz and K. F. Freed, 
{\em Macromolecules} 
{\bf 23}, 1519 (1990);
H. Tang and K.F. Freed,
{\it J. Chem. Phys.}
{\bf 94}, 1572 (1991);
M. Lifschitz and K.F. Freed,
{\it J. Chem. Phys.}
{\bf 98}, 8994 (1993);
K. F. Freed, 
{\it J. Chem. Phys.}
{\bf 103}, 3230 (1995);
K.W. Foremann and K.F. Freed,
{\it J. Chem. Phys.}
{\bf 106}, 7422 (1997).
\bibitem{schweizer} 
K. S. Schweizer and J. G. Curro, in 
{\em Advances in Chemical Physics},
Vol XCVIII, I. Prigogine and S. A. Rice (eds.), 
Wiley, New York(1997).
\bibitem{friederike1}
For a recent review see
F. Schmid, 
{\it Journal of physics: Condensed matter},
preprint 1998.
\bibitem{andreas2}
A. Werner, F. Schmid, M. M\"uller, and K. Binder,
{\it J. Chem. Phys.}
{\bf 107}, 8175 (1997).
\bibitem{ck}
I. Carmesin and K. Kremer,
{\it Macromolecules}
{\bf 21}, 2819 (1988);
{\it J. Phys. (France)}
{\bf 51}, 915 (1990).
\bibitem{binder}
K. Binder (edt.),
{\it Monte Carlo and Molecular Dynamics Simulations in Polymer Science},
Oxford University Press, Oxford (1995).
\bibitem{marcusr}
M. M\"uller and F. Schmid, in
{\it Annual Reviews of Computational Physics},
D. Stauffer edt.,
World Scientific, Singapore,
preprint 1998.
\bibitem{marcus3}
M. M\"uller and M. Schick,
{\it J. Chem. Phys.}
{\bf 105}, 8885 (1996).
\bibitem{gary}
M.D. Lacasse, G.S. Grest, and A.J. Levine,
{\it Phys. Rev. Lett.}
{\bf 80}, 309 (1998).
\bibitem{marcus1}
M. M\"uller, K. Binder, and W. Oed,
{\it J. Chem. Soc. Faraday Trans.}
{\bf 91}, 2369 (1995).
\bibitem{friederike2}
F. Schmid and M. M\"uller,
{\it Macromolecules}
{\bf 28}, 8639 (1995).
\bibitem{marcus2}
M. M\"uller and A. Werner,
{\it J. Chem. Phys.}
{\bf 107}, 10764 (1997).
\bibitem{pbhk}
W. Paul, K. Binder, D.W. Heermann, and K. Kremer,
{\it J. Phys. II (France)}
{\bf 1}, 37 (1991).
\bibitem{marcus4}
M. M\"uller and K. Binder,
{\em Macromolecules}
1998, in press.
\bibitem{kb}
K. Kremer and K. Binder,
{\it Comp. Phys. Rep.}
{\bf 7}, 261 (1988).
\bibitem{degennes}
P.G. de Gennes,
{\it Scaling Concepts in Polymer Physics},
Cornell University, Ithaca (1979).
\bibitem{marcus5}
M. M\"uller and K. Binder,
{\it Macromolecules}
{\bf 28}, 1825 (1995).
\bibitem{joerg}
J. Baschnagel and K. Binder,
{\em Physica} A {\bf 204}, 47 (1994).
\bibitem{marcus6}
M. M\"uller and W. Paul,
{\it J. Chem. Phys.}
{\bf 100}, 719 (1994).
\bibitem{broseta}
D. Broseta, G.H. Fredrickson, E. Helfand, and L. Leibler
{\it Macromolecules}
{\bf 23}, 132 (1990).
\bibitem{freed2}
H. Tang and K.F. Freed,
{\it J. Chem. Phys.}
{\bf 94}, 3183 (1991).
\bibitem{sb0}
A. Sariban and K. Binder,
{\it J. Chem. Phys.}
{\bf 86}, 5859 (1987);
{\it Macromolecules}
{\bf 21}, 711 (1988).
\bibitem{KP}
O. Kratky and G. Porod, 
{\it Rec. Trav. Chim.} 
{\bf 68}, 1106 (1949);
N. Saito, K. Takahashi, and Y. Yunoki, 
{\em J. Phys. Soc. Jpn.} 
{\bf 22}, 219.(1967).
\bibitem{mf}
D.C. Morse and G.H. Fredrickson,
{\it Phys. Rev. Lett.}
{\bf 73}, 3235 (1994).
\end{thebibliography}
\end{document}